\documentclass{pasa}%

\usepackage{graphicx}	
\usepackage{amsmath}	
\usepackage{amssymb}	

\usepackage{natbib} 
\usepackage[usenames,dvipsnames]{color}
\usepackage{verbatim}
\usepackage{multirow}
\usepackage{booktabs}
\usepackage{subfigure}
\usepackage{colortbl}

\usepackage[T1]{fontenc}
\usepackage{mathptmx}
\usepackage[tracking=smallcaps]{microtype}
\SetTracking[no ligatures = f]{encoding = *,shape = sc}{0}

\usepackage{listings}

\usepackage{bold-extra}
\usepackage{slantsc}
\usepackage{times}

\definecolor{verbgray}{gray}{0.9}
\definecolor{shadecolor}{rgb}{.9, .9, .9}
\definecolor{lightred}{rgb}{0.96,0.81,0.81}
\definecolor{verylightceleste}{rgb}{0.91, 0.93, 0.97}
\definecolor{lightceleste}{rgb}{0.75, 0.81, 0.98}
\definecolor{celeste}{rgb}{0.45, 0.58, 0.82}
\definecolor{white}{rgb}{1., 1., 1.}
\definecolor{darkblue}{rgb}{0.10, 0.35, 0.99}
\definecolor{darkblue2}{rgb}{0.10, 0.24, 0.59}
\definecolor{verydarkblue}{rgb}{0.05, 0.10, 0.24}
\definecolor{black}{rgb}{0., 0., 0.}
\definecolor{lightviolet}{rgb}{0.95, 0.89, 0.96}
\definecolor{violet}{rgb}{0.7, 0.28, 0.70}
\definecolor{darkviolet}{rgb}{0.42, 0.24, 0.41}
\definecolor{green}{rgb}{0.18, 0.41, 0.02}
\definecolor{lightgreen}{rgb}{0.62, 0.75, 0.43}
\definecolor{darkgreen}{rgb}{0.27, 0.34, 0.16}
\definecolor{verylightgreen}{rgb}{0.298039,0.952941,0.345098}
\definecolor{verylightgreen2}{rgb}{0.745098,1,0.654902}
\definecolor{yellow}{rgb}{1, 1, 0}
\definecolor{darkyellow2}{rgb}{1, 0.890196,0.427451}
\definecolor{darkyellow}{rgb}{0.87, 0.87, 0.03}
\definecolor{darkyellow3}{rgb}{1,0.933333,0.654902}
\definecolor{gray}{rgb}{0.14, 0.14, 0.15}
\definecolor{lightgray}{rgb}{0.70, 0.72, 0.74}
\definecolor{silver}{rgb}{0.91, 0.91, 0.91}
\definecolor{Blueb}{cmyk}{0.2,0,0,0}
\definecolor{lightorange}{rgb}{1,0.886275,0.619608}

\newcommand{\caesar}{\textnormal{\textsc{Caesar}} }
\newcommand{\um}[1]{\ \mathrm{#1}}

\newcommand{\hii}{H\textsc{ii} region}

\lstnewenvironment{codebkg}{%
  \lstset{
    escapechar=\&,
    basicstyle=\ttfamily\footnotesize,
    backgroundcolor=\color{verbgray},
    frame=single,
    framerule=0pt,
    columns=fullflexible
  }
 }
 {}

\title[Caesar testing]{\caesar source finder: recent developments and testing}

\author[S. Riggi et al.]{
S. Riggi$^1$\thanks{E-mail: simone.riggi@inaf.it} , %
F. Vitello$^1$, %
U. Becciani$^1$, %
C. Buemi$^1$, %
F. Bufano$^1$, %
A. Calanducci$^1$, %
F. Cavallaro$^1$, %
A. Costa$^1$,\\%
A. Ingallinera$^1$, %
P. Leto$^1$, %
S. Loru$^1$, %
R.P. Norris$^{2,3}$, %
F. Schillir\`{o}$^1$, %
E. Sciacca$^1$, %
C. Trigilio$^1$, %
G. Umana$^1$%
\vspace{0.2cm}
\affil{$^1$INAF-Osservatorio Astrofisico di Catania, Via Santa Sofia 78, 95123 Catania, Italy}%
\affil{$^2$CSIRO, P.O. Box 76, Epping, NSW 1710, Australia}
\affil{$^3$Western Sydney University, Penrith, NSW, Australia}

}%

\jid{PASA}
\doi{\url{https://doi.org/10.1017/pasa.2019.29}}
\jyear{\the\year}

\usepackage{aas_macros}
\usepackage{hyperref} 
\hypersetup{colorlinks,citecolor=blue,linkcolor=blue,urlcolor=blue}

\begin{document}

\begin{frontmatter}
\maketitle

\begin{abstract}
A new era in radioastronomy will begin with the upcoming large-scale surveys planned at the Australian Square Kilometre Array Pathfinder (ASKAP). ASKAP started its Early Science program in October 2017 and several target fields were observed during the array commissioning phase. 
The \textsc{Scorpio} field was the first observed in the Galactic Plane in Band 1 (792-1032 MHz) using 15 commissioned antennas. The achieved sensitivity and large field of view already allow to discover new sources and survey thousands of existing ones with improved precision with respect to previous surveys. 
Data analysis is currently ongoing to deliver the first source catalogue. Given the increased scale of the data, source extraction and characterization, even in this Early Science phase, have to be carried out in a mostly automated way. This process presents significant challenges due to the presence of extended objects and diffuse emission close to the Galactic Plane.\\
In this context we have extended and optimized a novel source finding tool, named \textsc{Caesar}, to allow extraction of both compact and extended sources from radio maps. A number of developments have been done driven by the analysis of the \textsc{Scorpio} map and in view of the future ASKAP Galactic Plane survey. The main goals are the improvement of algorithm performances and scalability as well as of software maintainability and usability within the radio community. In this paper we present the current status of \textsc{Caesar} and report a first systematic characterization of its performance for both compact and extended sources using simulated maps. Future prospects are discussed in light of the obtained results.
\end{abstract}

\end{frontmatter}

\section{INTRODUCTION }
\label{sec:intro}
The Square Kilometer Array (SKA) precursor era has finally come with the opening of
the Australian SKA Pathfinder (ASKAP) Early Science program in October 2017. While
the deployment phase is still ongoing, a number of target fields are being observed
with the commissioned antennas to demonstrate ASKAP scientific capabilities, validate 
imaging pipeline and facilitate the development of analysis techniques in view of the 
operations with the full 36-antenna array. In particular, the \textsc{Scorpio} survey field 
($\sim$40 square degrees in size, centered on $l=343.5^\circ$, $b=0.75^\circ$) was observed 
in January 2018 in ASKAP Band 1 ($912\um{MHz}$) with 15 antennas. Details on the observation strategy and data reduction will be presented in a forthcoming paper.\\The \textsc{Scorpio} survey \citep{Umana2015}, started in 2011 with a pilot program conducted with the Australian Telescope Compact Array (ATCA), has a clear scientific goal, which is the study and characterization of different types of Galactic radio sources, from stars to circumstellar regions (\hii s, planetary nebulae, luminous blue variables, Wolf--Rayet stars) and stellar relics (e.g. supernova remnants).     
Besides its scientific goals, it represents an important test bench for imaging and analysis techniques in the Galactic Plane in view of the upcoming ASKAP Evolutionary Map of the Universe (EMU) survey \citep{Norris2011}, planned to start at the end of 2019.\\In this context the accuracy of source finding algorithms is still a concern considering that the size of the EMU survey in terms of surveyed area and number of expected sources will severely limit a manual intervention on the source cataloguing process.

\begin{figure*}[!ht]
\begin{center}
\includegraphics[scale=0.4]{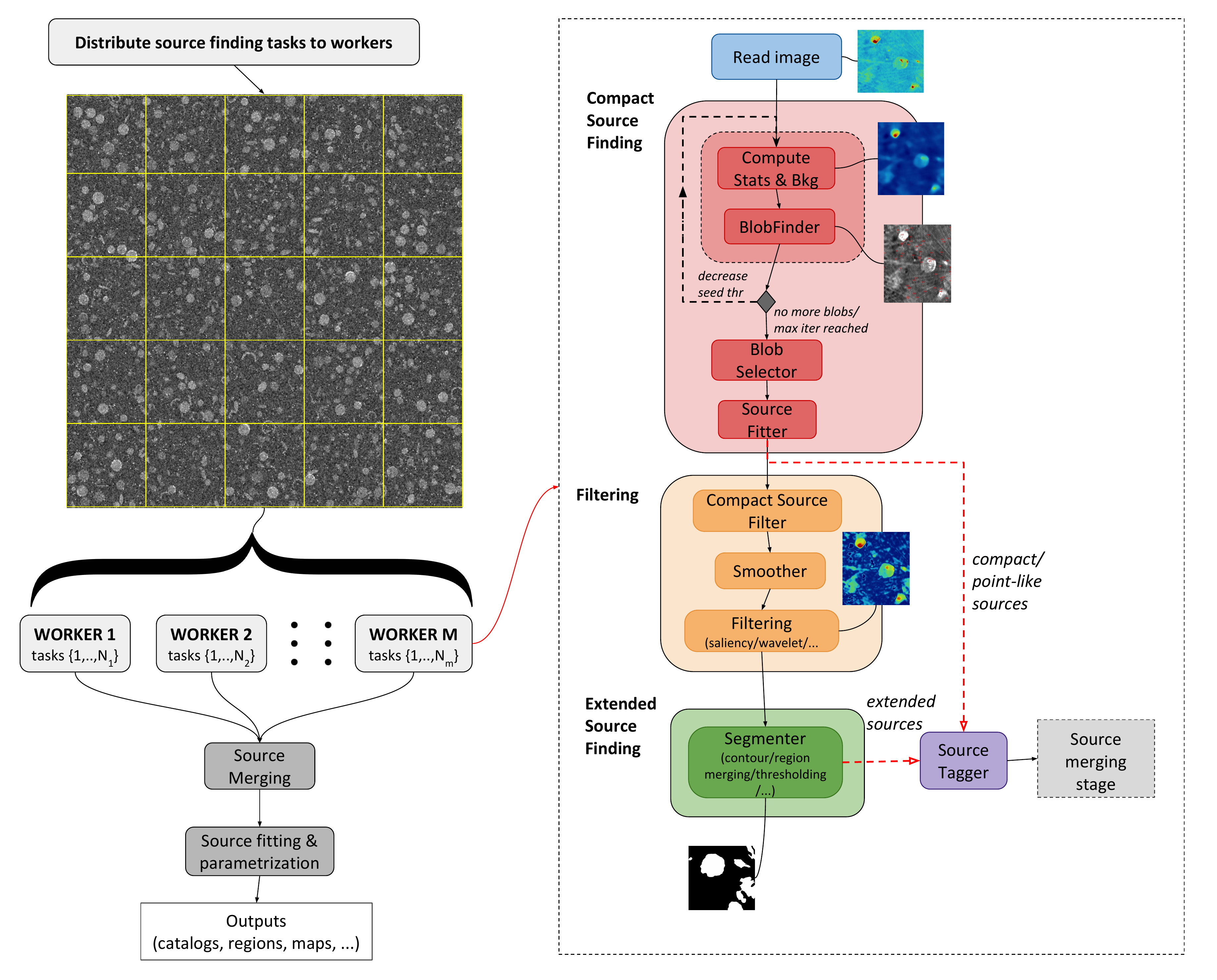}
\caption{A schema of \caesar source finding pipeline. See text for a description of pipeline stages. Compact and extended source finding stages are described in Sections~\ref{sec:compactsources} and ~\ref{sec:extsourceextraction} respectively. Filtering and source merging stages are described in Section~\ref{sec:residualimg} and ~\ref{sec:sourcemerging}.}%
\label{fig:pipeline}
\end{center}
\end{figure*}

Significant efforts have been spent within the ASKAP EMU Collaboration\footnote{\url{http://askap.pbworks.com/TeamMembers}} to systematically compare different source finders, evaluating their performances on simulated data samples \citep{Hopkins2015}. These analysis pointed out strong and weak features of the tested finders, triggering new developments in specific areas, such as source deblending and fitting (e.g. see \cite{Hancock2018} and \cite{Carbone2018} for recent works). Existing works, however, concentrate on compact sources, and well-known source finders, such as Aegean \citep{Hancock2012}, PyBDSF \citep{Mohan2015} and \textsc{Blobcat} \citep{Hales2012}, have been shown not to perform well on extended sources, revealing the need for further developments in the characterization of complex extended sources and for a systematic testing with simulations.\\The \caesar source finder \citep{Riggi2016} was developed to overcome some of these issues and to provide missing features, particularly for the analysis of radio maps in the Galactic Plane.\\This paper has multiple goals. Firstly we report the status of \caesar and recent developments made since \cite{Riggi2016} in Section ~\ref{sec:caesardev}. Secondly, we resume the ongoing efforts to systematically characterize and evaluate the source detection accuracy and computational performances with simulated data. In Section~\ref{sec:simulations} we describe the simulated data sample produced to test \caesar performances. The analysis carried out on simulated data are reported in Section~\ref{sec:analysis}. Performance results (completeness, reliability, etc) obtained on both compact and extended sources are presented and discussed. In Section~\ref{sec:performances} we analyze the computational performance (cpu and memory usage, scalability, etc) obtained in multithreaded and parallel runs performed over a test computing infrastructure. Finally, in Section~\ref{sec:summary} we discuss the \caesar roadmap and further analysis to be carried out, taking into consideration the results obtained in this paper.\\%
This work constitutes also part of the ongoing analysis for the preparation of ASKAP \textsc{Scorpio} Early Science source catalog. Besides the \textsc{Scorpio} and ASKAP EMU Galactic program, this work is well suited in the context of SKA \emph{OurGalaxy} key science project and the European SKA Regional Data Center (ESDC) design\footnote{Details on the SKA ESDC design and AENEAS EU H2020 project available at \url{https://www.aeneas2020.eu/}}. Indeed, it is anticipated that the SKA Science Data Processor (SDP) will invest limited resources for the development, optimization and testing of science algorithms particularly for the Galactic science \citep{Hollitt2016}. These activities have therefore to be largely lead in synergy by science and ESDC working groups.

\begin{figure*}[!ht]
\begin{center}
\subtable{\includegraphics[scale=0.39]{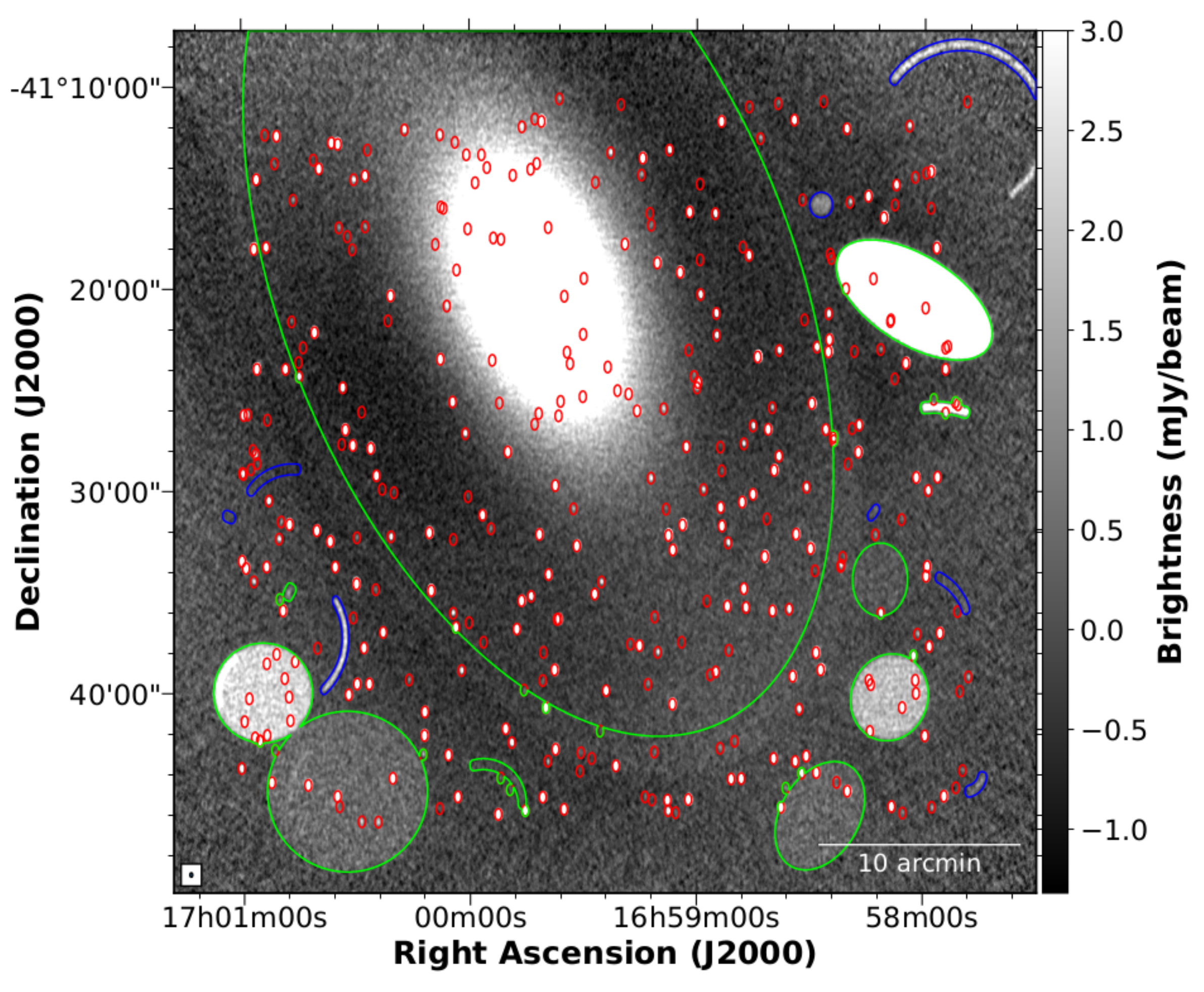}}
\hspace{-0.1cm}%
\subtable{\includegraphics[scale=0.39]{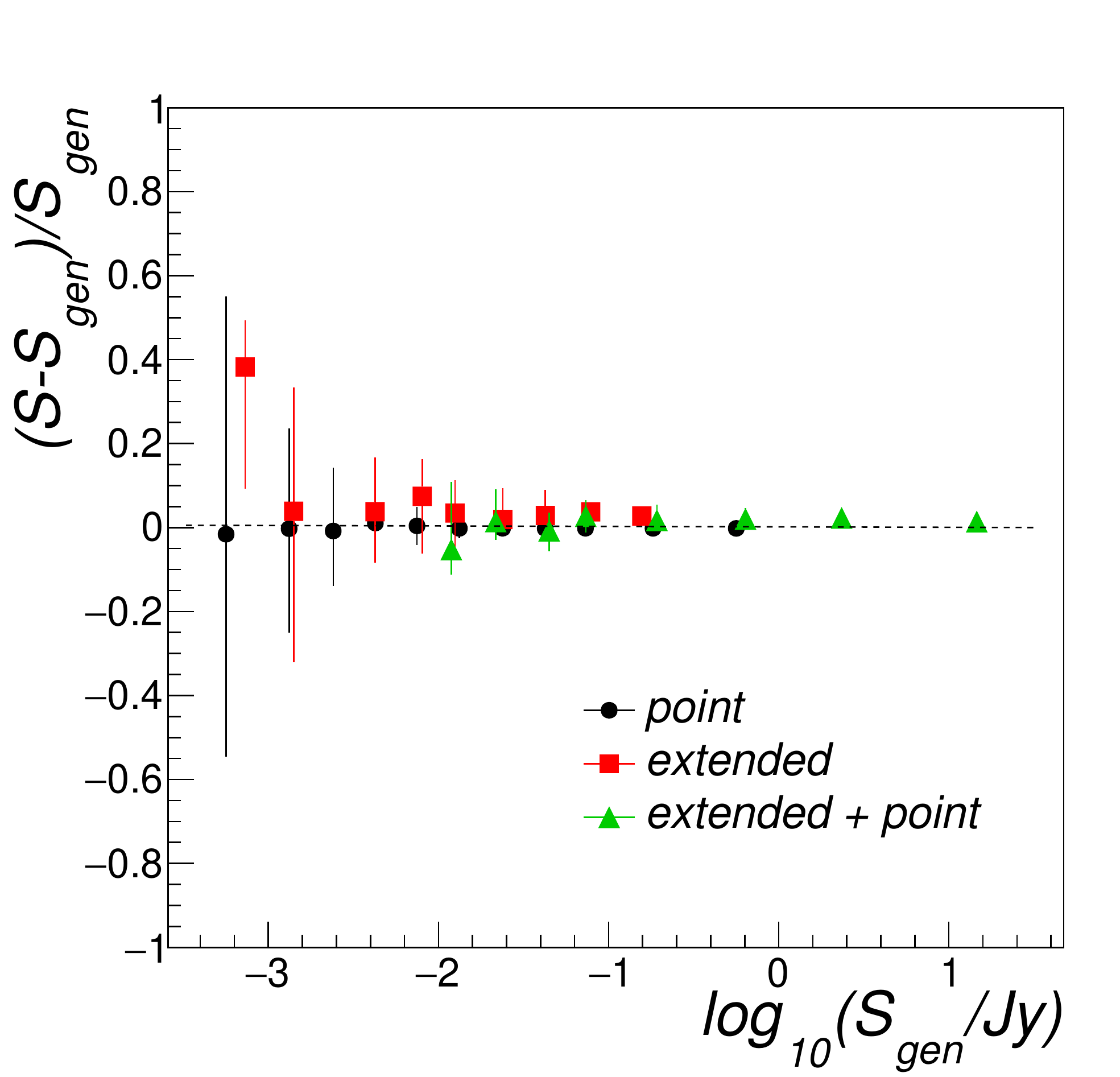}}
\caption{Left: Sample simulated map in mJy/beam units with convolved source contours superimposed (red=point-like, blue=extended, green=extended+point-like); Right: Imaging flux accuracy for point sources (black dots), extended sources with nested point sources (green triangles), isolated extended sources (red squares) obtained on the simulated data set. Each dot represents the median of the pull distribution ($S$-$S_{gen}$)/$S_{gen}$ in $log_{10}S_{gen}$ bins, being $S_{gen}$ the generated source flux density (after convolution with the synthesis beam as described in the text) and $S$ the imaged source flux density. Error bars are the interquartile range of the pull distribution.}
\label{fig:imagingperf}
\end{center}
\end{figure*}

\section{\caesar: status and recent developments}
\label{sec:caesardev}
\caesar \citep{Riggi2016} is a C++ tool for extraction of compact and extended sources from astronomical images developed in the context of the \textsc{Scorpio} project and ASKAP EMU survey. It is based on 3rd-party libraries and software frameworks, among them ROOT \citep{Brun1997}, OpenCV \citep{Bradski2000}, MPI library\footnote{\url{www.open-mpi.org}}.\\ 
A number of improvements and developments have been done in distinct areas since the original work \citep{Riggi2016}, summarized below:
\begin{itemize}
\item \textbf{Code refactoring}: the software code was updated and reorganized to improve modularity and maintainability and to lower memory demand. Dependency on some of the external libraries (R, OpenMP, MPI) was made optional.
\item \textbf{Algorithm optimization and speed-up}: Recurrent tasks, including image reading, statistics and background estimation, image filtering, were optimized and parallelized using OpenMP\footnote{\url{www.openmp.org}} directives, whenever a benefit in speed-up was identified. For example, computation of image median estimators was optimized to improve the original time complexity from O(N log(N)) to O(N). Statistical moments (up to 4th order) are computed using online parallel formulas (e.g. while reading and filling the image in different threads). Benchmark tests were carried out against corresponding python implementations (mostly based on python numpy module) and a speed up $\sim$12 was found on sample images of size 32000$\times$32000 pixels.\\Newer parallel algorithms available in the standard C++ library (e.g. parallel \textit{nth\_element}) were also tested and benchmarked against the corresponding non parallel version. No significant improvements were found in this case. 
\item \textbf{Distributed processing}: A parallel MPI-based version of the source finder application was implemented to support distributed processing of large maps on different computing nodes. Multi-thread processing per node, based on OpenMP, is also available and configurable. A serializer, based on the Google Protocol Buffer library\footnote{\url{https://developers.google.com/protocol-buffers/}}, was added to allow serialization/deserialization of \textsc{Caesar} objects when exchanging data across computing nodes.
\item \textbf{Logging}: Custom logging macros were added to all components and applications using log4cxx library\footnote{\url{https://logging.apache.org/log4cxx/index.html}}. Logging levels can be customized from a configuration file or programmatically.
\item \textbf{Algorithm improvements and extensions}: Compact source finding was improved in different aspects with respect to previous version. Details are reported in the following sections. Additional applications, besides source finding, were added to ease post-processing tasks, such as source cross-matching and analysis.
\item \textbf{Distribution and usability}: Efforts have been made to make \caesar publicly available at \url{https://github.com/SKA-INAF/caesar.git}, portable and usable in different systems with limited effort. To this aim we provide recipe files to build and run \caesar applications in a Singularity\footnote{\url{https://singularity.lbl.gov/}} container. Details on how to use \caesar are given in the online documentation at \url{https://caesar-doc.readthedocs.io/en/latest/}.
\end{itemize}

\subsection{Processing pipeline}
A schema of the processing pipeline is shown in Fig.~\ref{fig:pipeline}.
The input image is partitioned into sub-images or tiles according to configurable parameters (e.g. tile size, overlap, etc). Tiles are then distributed among available workers for processing. Each processor, thus, effectively reads and keeps in memory only a portion of the input image, corresponding to the assigned tiles.
Each worker executes the source finding pipeline on the assigned tiles in sequence. This includes a series of steps, shown in Fig.~\ref{fig:pipeline} for one representative processor and tile, summarized below:

\begin{enumerate}
\item Extract compact sources from input tile through the following stages (see section~\ref{sec:compactsources} for more details):
\begin{enumerate}
  \item Compute image statistic estimators, global and local background, noise and significance maps;
  \item Iteratively extract blobs using significance map and nested blobs using a blob-sensitive filter map;
  \item Reject anomalous blobs and promote blobs to source candidates, tagging them as compact or point-like;
  \item Fit and parametrize source candidates present in the tile and tag sources located at the borders 
\end{enumerate}
\item Compute a residual map, obtained from input map by applying one or more filters (smoothing, filters to enhance diffuse emission) after removal of compact bright sources (see section~\ref{sec:residualimg});
\item Extract extended sources from residual map according to the selected algorithm and tag them accordingly (see section~\ref{sec:extsourceextraction});
\item Merge adjacent and overlapping compact and extended sources found in the tile (see section~\ref{sec:sourcemerging});
\end{enumerate}
A master processor aggregates sources extracted by other workers, merging them if overlapping or adjacent at tile borders. Source fitting and parametrization is finally performed on merged sources (if any) and outputs (e.g. catalogs with sources and fitted components, regions, etc) are delivered as final results.\\Details on the computing stages and specific algorithms can be found in \cite{Riggi2016}. In the following sections we limit the discussion on the improvements made in the new version, mainly relative to compact source extraction.

\subsubsection{Compact source extraction}
\label{sec:compactsources}
Compact source extraction is based on four stages:

\begin{enumerate}
\item \textit{Blob search}: 
Blobs are extracted from the input map with a flood-fill algorithm using a pixel significance detection threshold $Z_{thr,d}$ (usually equal to 5) and a lower aggregation threshold $Z_{thr,m}$ (usually equal to 2.5). Pixel significance level $Z$ is computed as: 
\begin{equation}
Z= \frac{S-\mu_{bkg}}{\sigma_{rms}}    
\end{equation}
where $S$ is the pixel flux and $\mu_{bkg}$, $\sigma_{rms}$ are the estimated background level and noise rms, respectively.\\Blob extraction can now be performed using an iterative procedure in which the background and noise maps are re-computed at each iteration without pixels belonging to sources extracted in the previous iterations. Detection thresholds can be progressively lowered by a configurable amount $\Delta Z$ (0.5 by default) at each $j$-th step until a maximum number of iterations is reached:

\begin{equation*}
Z_{thr,d}^{(j)}=Z_{thr,d}^{(0)}-j\times\Delta Z 
\end{equation*}

\item \textit{Nested blob search}: 
A blob detector algorithm can be applied on the input map to search for "nested" (or "child") blobs inside the "primary" (or "mother") blobs extracted in the previous step. Nested blobs are used in the image residual and source fitting stages (described in the following paragraphs).  
When enabled, the algorithm proceeds as follows:
\begin{itemize}
\item A primary blob mask is obtained using blobs detected with the flood-fill approach;
\item A blob-sensitive filter is applied to the input map and blobs are searched in the resulting filtered map using flood-fill method around detected peaks above a specified significance threshold (typically equal to 5). Extracted blobs are then used to build a secondary blob mask;
\item The secondary blob mask is cross-matched against the primary one to extract nested blobs and associate them to primary blobs.
\end{itemize}
Two alternative blob-sensitive filter models are provided (multiscale Laplacian of Gaussian (LoG), elliptical gaussian) with customizable kernel size and scale parameters (first/last scale, scale increment). The first method can be computationally demanding if the chosen kernel size is large (e.g. say above 9-11 pixels) and several scales are requested. 
The second approach is relatively fast as it employs only one scale, i.e. the elliptical beam of the input map.\\
Nested blob search can be disabled if not explicitly needed (typically in the absence of extended sources) or, alternatively, customized. For example, nested blobs can be searched only on sources tagged as extended, i.e. exceeding a certain area-to-beam threshold factor (usually set to 10-20).

\begin{figure*}[!ht]
\begin{center}
\subtable[Completeness]{\includegraphics[scale=0.4]{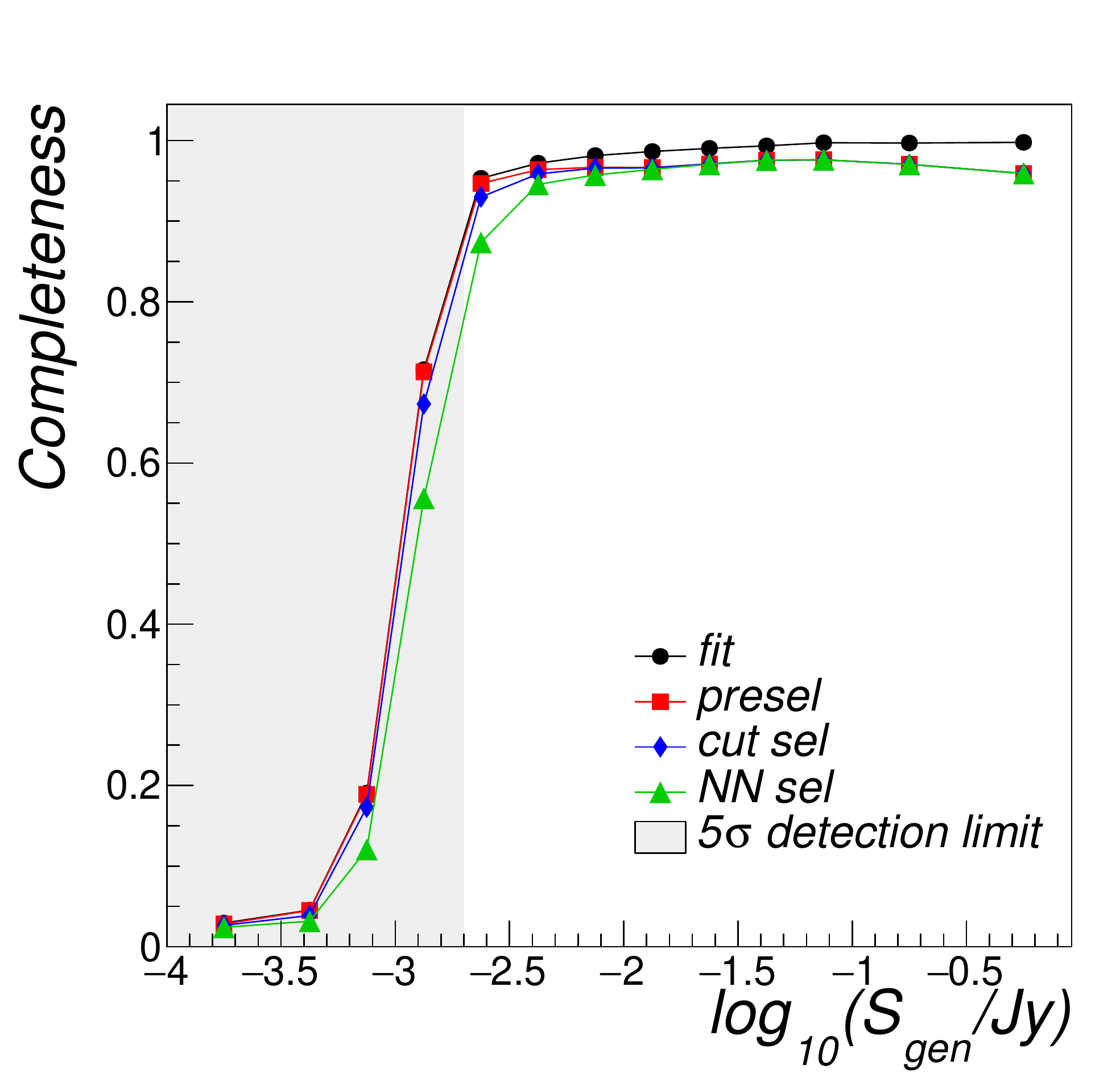}}%
\subtable[Reliability]{\includegraphics[scale=0.4]{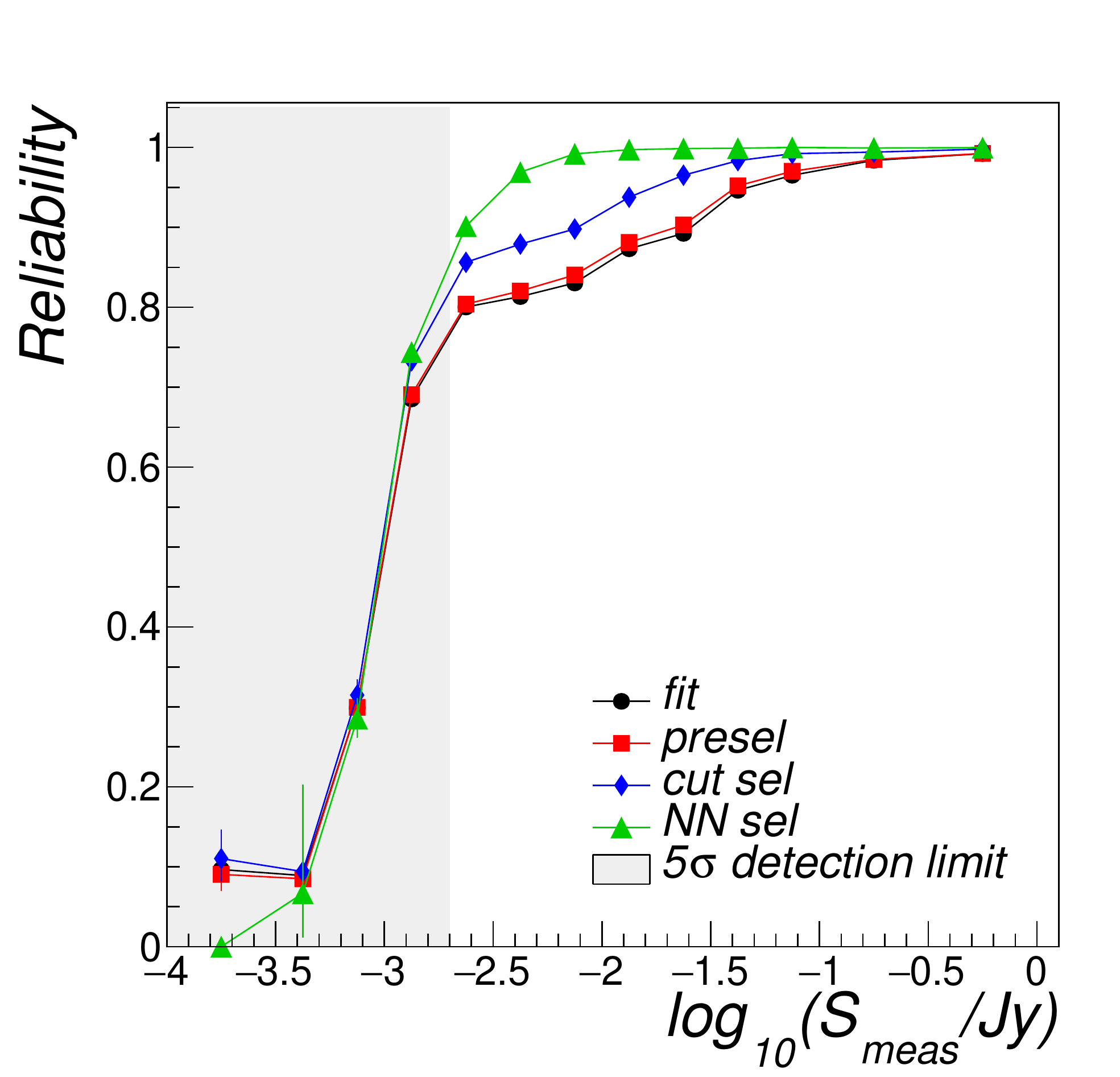}}%
\caption{Left: Compact source detection efficiency as a function of the generated source flux density for four different source selections (described in the text): fit converged (black dots), preselection cuts (red squares), preselection + cut selection (blue diamonds), preselection + neural network selection (green triangles); Right: Compact source detection reliability as a function of the measured source flux density for four different source selections (described in the text): fit converged (black dots), preselection cuts (red squares), preselection + cut selection (blue diamonds), preselection + neural network selection (green triangles).}
\label{fig:compactsourcecompleteness}
\end{center}
\end{figure*}

\item \textit{Blob selection}:
Extracted blobs are selected using simple morphological parameters (blob area-beam ratio, roundness, elongation, bounding box, etc) to tag candidate point-like sources and exclude anomalous blobs with elongated shapes, most likely due to imaging artifacts.

\item \textit{Source deblending and fitting}:
Source fitting is performed by workers on sources that are not located at the tile borders and by the master processor on merged edge sources. The adopted fitting procedure depends on the detected source size. For extended sources, i.e. above a configured area-to-beam ratio, only nested blobs (if any) are individually fitted. Compact sources (e.g. nested or not and below the area-to-beam ratio threshold) are fitted with a mixture of $M$ gaussian components, plus a background offset parameter $S_0$. The following $\chi^{2}$ is minimized with respect to $(M+1)$ fitting parameters $\boldsymbol\Theta$=\{$S_0$,$\boldsymbol\Theta_{1}$,\dots,$\boldsymbol\Theta_{M}$\}, where $\Theta_k$=\{$\bar{x_k}$, $\bar{y}_k$, $\sigma_{x_{k}}$, $\sigma_{y_{k}}$, $\theta_k$\}:
\begin{equation}
\chi^{2}= \sum_{i=1}^{N} \frac{[S_{i}(x_i,y_i)-\hat{S}_{i}(x_i,y_i;\Theta)]^{2}}{\sigma_{i}^{2}}\;\;%
\end{equation}
where:
\begin{equation}
\hat{S}(x_i,y_i,\Theta)= S_{0} + \sum_{k=1}^{M}f_{k}(x_i,y_i;\boldsymbol\Theta_{k})
\end{equation}
\begin{equation}
\begin{split}
f_{k}(x_i,y_i,\boldsymbol\Theta_k)=& A_{k}\exp[-a_{k}(x_i-\bar{x}_{k})^{2}-\\%
&- b_{k}(x_i-\bar{x}_{k})(y_i-\bar{y}_{k})-\\%
&- c_{k}(y_i-\bar{y}_{k})^{2}]
\end{split}
\end{equation}
\begin{equation}
a_k= \frac{\cos^{2}(\theta_k)}{2\sigma_{x_k}^{2}} + \frac{\sin^{2}(\theta_k)}{2\sigma_{y_k}^{2}}%
\end{equation}
\begin{equation}
b_k= \frac{\sin(2\theta_k)}{2\sigma_{x_k}^{2}} - \frac{\sin(2\theta_k)}{2\sigma_{y_k}^{2}}%
\end{equation}
\begin{equation}
c_k= \frac{\sin^{2}(\theta_k)}{2\sigma_{x_k}^{2}} + \frac{\cos^{2}(\theta_k)}{2\sigma_{y_k}^{2}}%
\end{equation}

with $N$ number of source pixels, $S_{i}$ and $\hat{S}_{i}$ the data and the predicted flux of the $i$-th source pixel respectively, $\sigma_{i}^{2}$ variance of the measurements. We assumed $\sigma_{i}$ equal to the estimated noise averaged over fitted source pixels. $A_{k}$ denotes the peak brightness of the $k$-th fitted component.\\
Total source flux density $I$ is computed as:
\begin{equation}
I=\sum_{k=1}^{M}I_{k},\;\;\;\;\;\;\;I_{k}=2\pi A_{k}\sigma_{x_k}\sigma_{y_k}
\end{equation}
with $I_k$ flux density of the $k$-th component. The flux density error $\delta I$ is computed by error propagation:
\begin{equation}
\delta I=\sqrt{\mathbf{D}\boldsymbol\Sigma\mathbf{D^{T}}},\;\;\;\;\;\;\;\;\mathbf{D}=\frac{\partial I}{\partial \boldsymbol\Theta}
\end{equation}
where $\mathbf{D}$ is the derivative matrix of flux density with respect to fit parameters $\boldsymbol\Theta$ and $\Sigma$ is the fit parameter covariance matrix.\\$\chi^{2}$ numerical minimization is performed with the \textsc{Root} minimizer libraries. Different minimizers\footnote{For multithreaded fitting Minuit2 has to be used as the other minimizers are not thread-safe} (e.g. \textit{Minuit} \citep{James1972}, \textit{Minuit2} \citep{Hatlo2005}, \textit{RMinimizer}) and minimization algorithms (e.g. \textit{Migrad}, \textit{Simplex}, \textit{BFGS}) are available to the user, all of them providing estimated errors on the fitted parameters as well as the fit parameter covariance matrix $\Sigma$. 
\\The approach followed to determine the optimal starting number of fitted components and relative parameters is usually denoted as the deblending process. Details are provided in Appendix~\ref{appendix:deblending}.\\All model parameters can be kept fixed or left free to vary in the fit and limits can be applied around parameter starting values. To guide fit convergence, the fit procedure is first performed with some parameters fixed (e.g. offset, component amplitudes) to the initial values. Fixed parameters are released afterwards and a full fit is performed. If one or more fitted parameters are found close or at the specified limits the fit procedure can be iteratively repeated, progressively enlarging the parameter range, until no more parameters are found at limits or a maximum number of retry iterations is exceeded. If the fit does not converge it can be repeated by progressively removing fainter fit components until convergence or until no more components are left.
\end{enumerate}

\begin{figure}[!ht]
\begin{center}
\subtable{\includegraphics[scale=0.36]{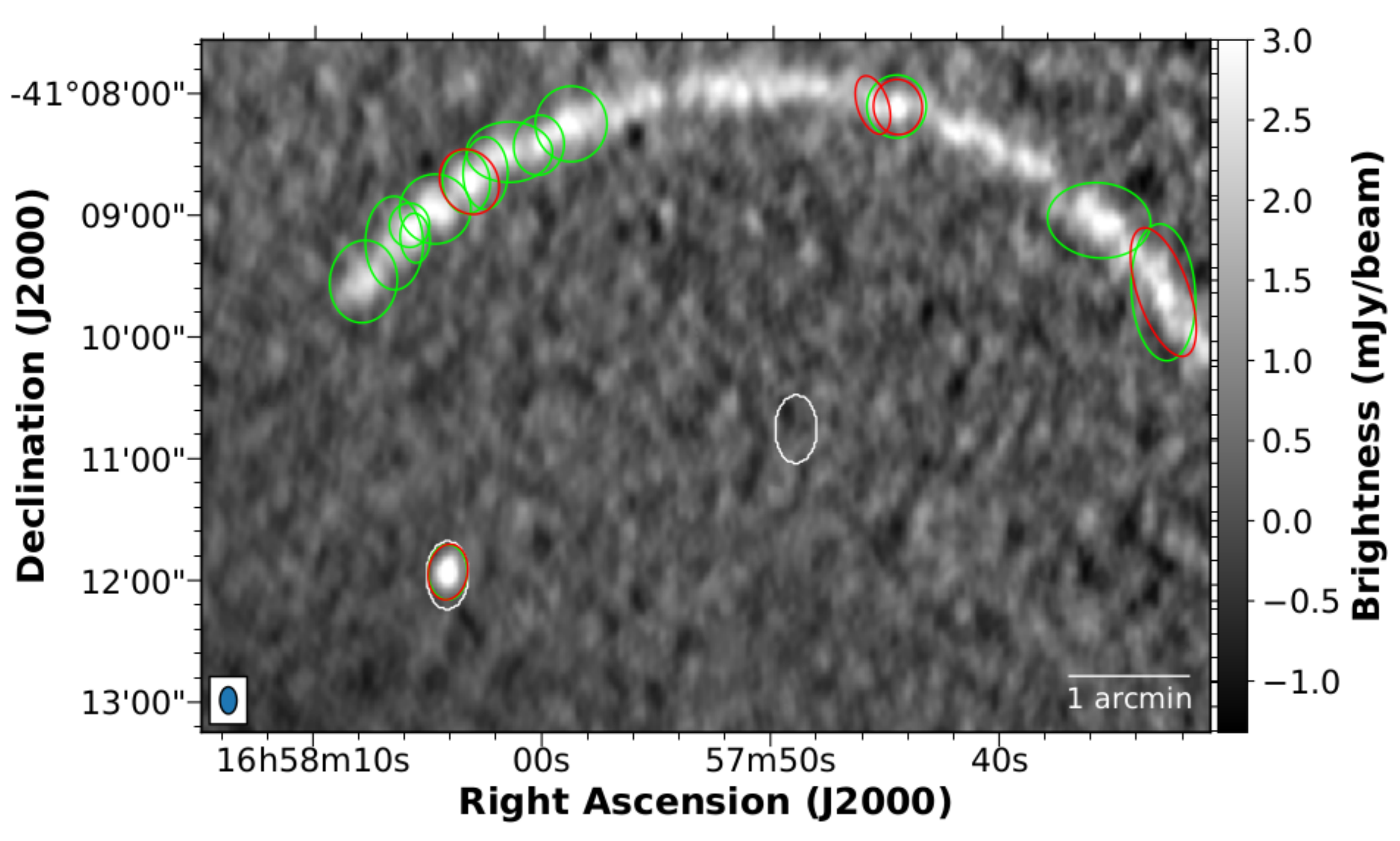}}%
\hspace{-0.0cm}
\subtable{\includegraphics[scale=0.34]{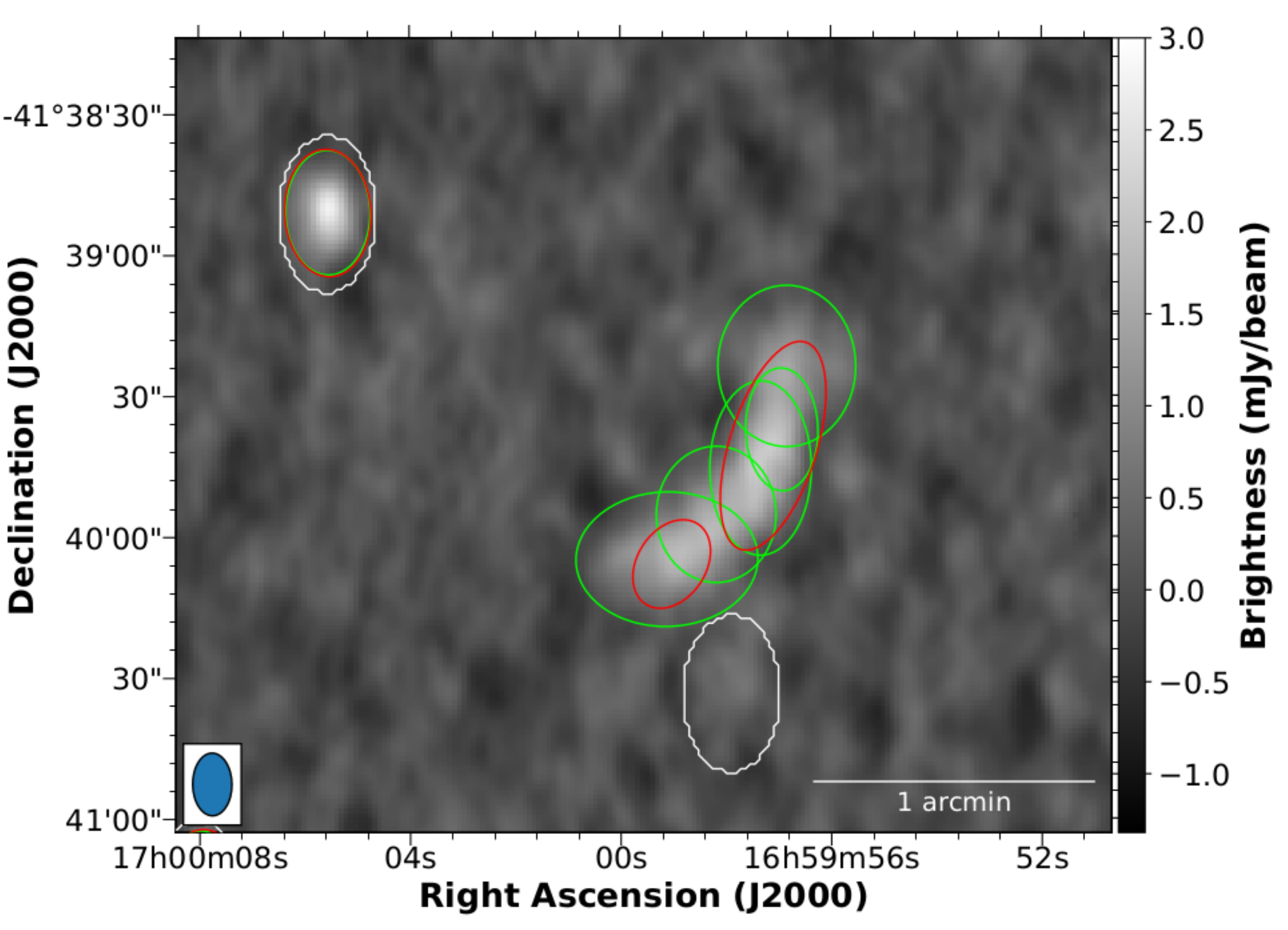}}%
\caption{Sample false compact sources detected by \textsc{Caesar} in simulated maps (red ellipses). Green ellipses represent sources detected by the \textsc{Aegean} source finder, while white ellipses represent generated point sources.}%
\label{fig:samplefalsesource}
\end{center}
\end{figure}

\subsubsection{Residual image and filtering}
\label{sec:residualimg}
Algorithms to extract faint extended sources are almost ineffective in presence of very bright point sources and noise artifacts in the field. For these reasons, the search is carried out on a residual image in which sources with peak flux above a configurable significance threshold with respect to the background (usually equal to 10) are removed from the map. Subtraction can be done in two alternative ways. The first method simply replaces all pixels belonging to bright sources with the estimated background\footnote{The algorithm uses a dilation filter to replace also pixels surrounding the source according to a configurable kernel size.}. The advantage of this approach, proposed in \cite{Peracaula2011}, is that it can be performed with only background information computed. A second, more refined, method subtracts the fitted model of bright sources from the input map. This, on the other hand, requires fit information to be available and accurate enough for the subtraction to be effective.\\
A series of filters can be applied to the residual image to limit the impact of small-scale artifacts and enhance the faint diffuse emission. A guided or gaussian smoothing filter is employed in \textsc{Caesar} for the former scope, while a Wavelet transform or saliency filter (see \cite{Riggi2016}) can be finally applied to produce the optimal input map for extended source search.

\begin{figure*}[!ht]
\begin{center}
\subtable{\includegraphics[scale=0.34]{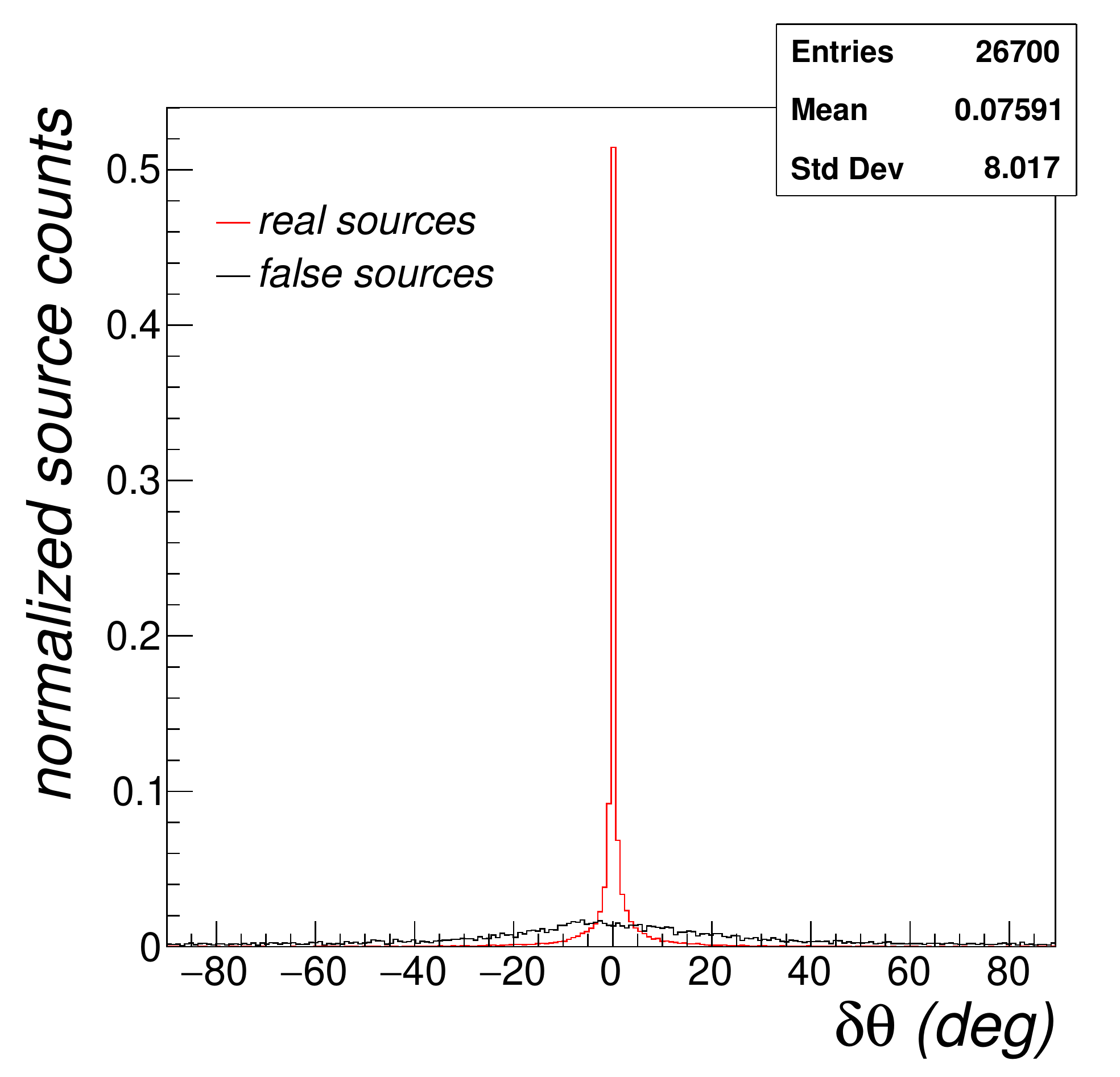}}%
\hspace{-0.1cm}%
\subtable{\includegraphics[scale=0.34]{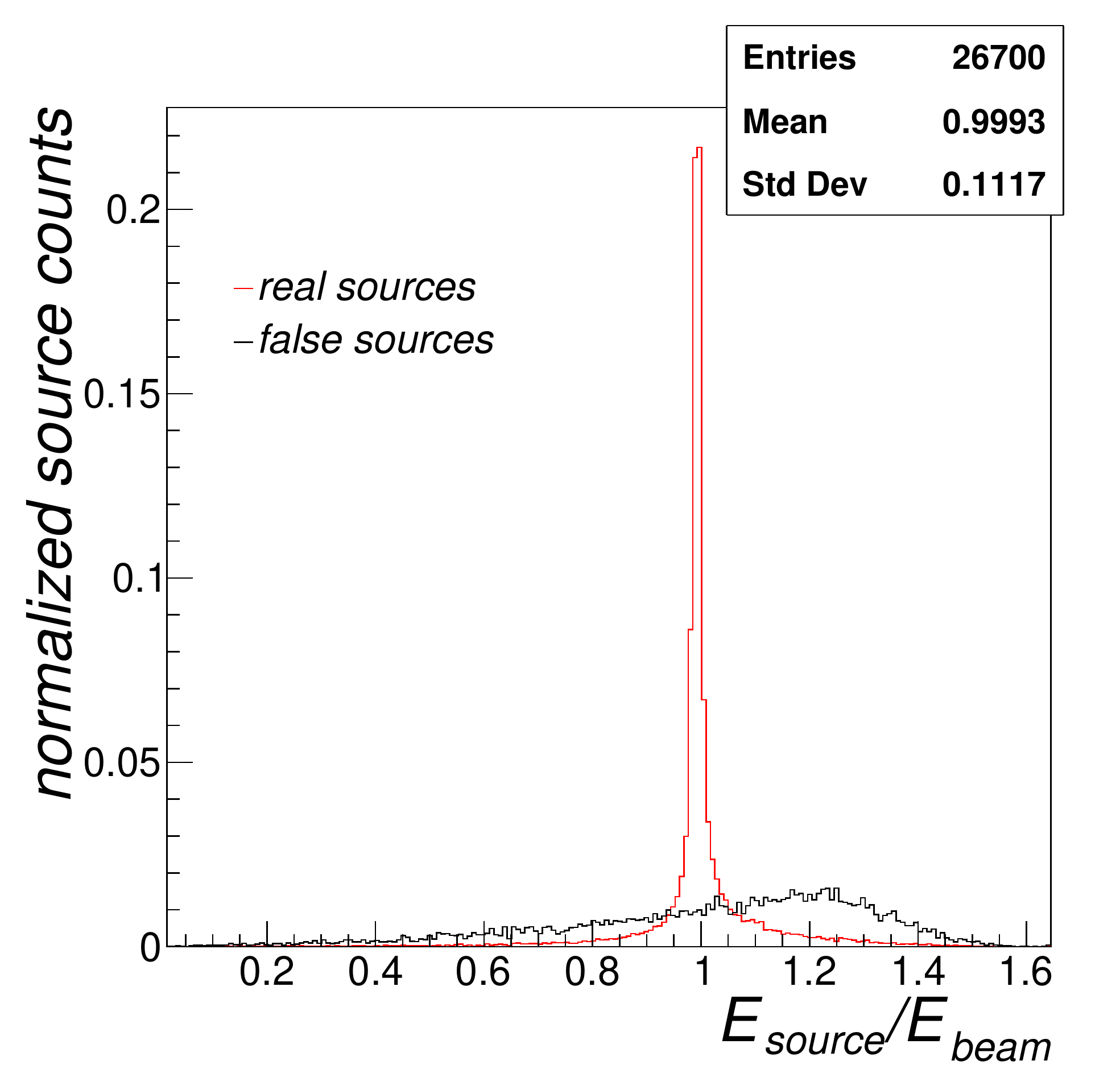}}%
\hspace{-0.1cm}%
\subtable{\includegraphics[scale=0.34]{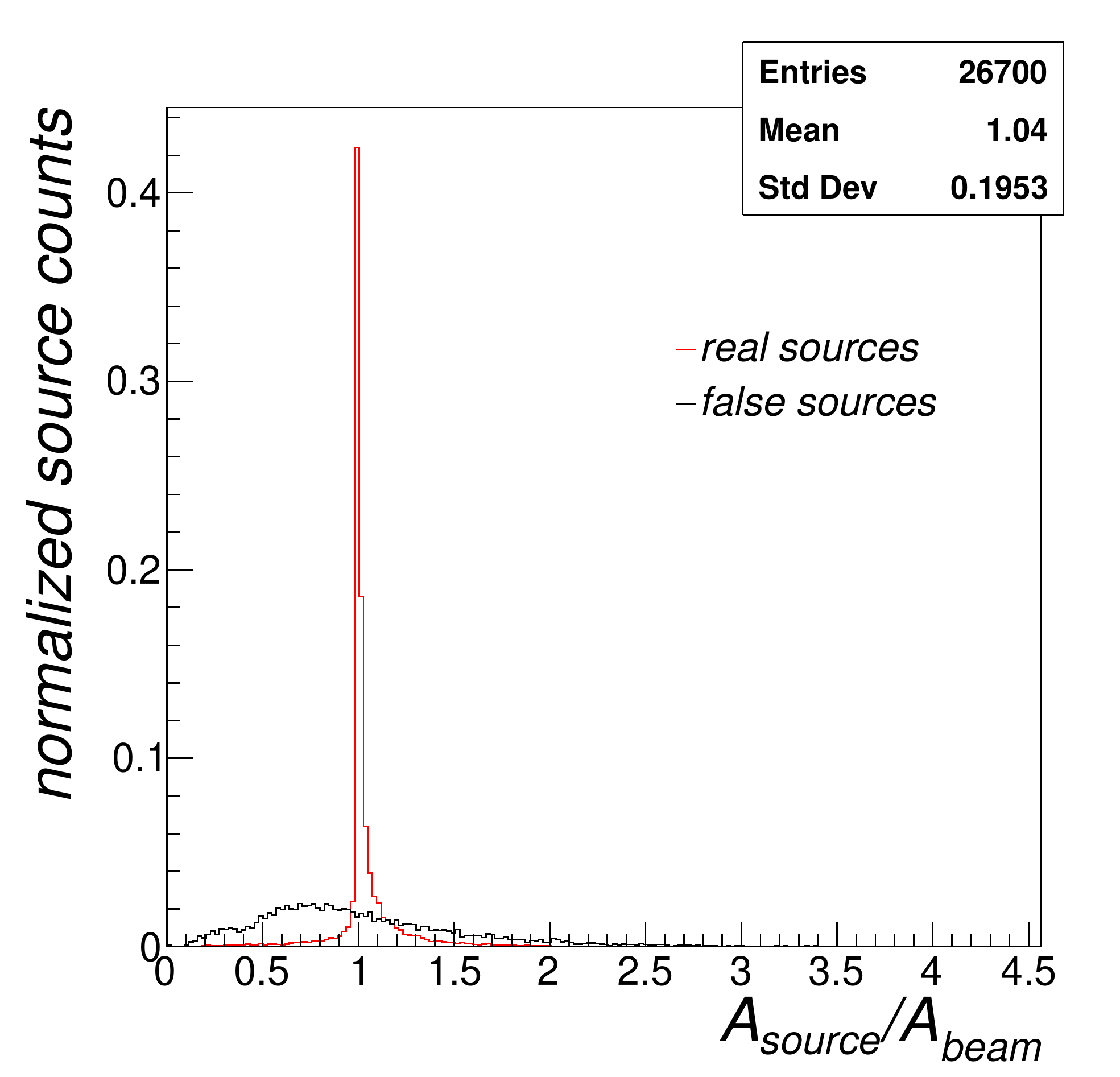}}%
\caption{Distribution of the classification parameters for real (red histogram) and false sources (black histogram). $\delta\theta$ (upper panel) represents the rotation angle (in degrees) of source fitted ellipse with respect to the beam ellipse. $E_{source}$/$E_{beam}$ (middle panel) represents the ratio between the source fitted ellipse eccentricity and the beam ellipse eccentricity. $A_{source}$/$A_{beam}$ (bottom panel) represents the ratio between the source fitted ellipse area and the beam area. Histograms are normalized to unit area with normalized counts reported in the y-axis.}
\label{fig:truefalseparhisto}
\end{center}
\end{figure*}

\subsubsection{Extended source extraction}
\label{sec:extsourceextraction}
Four classes of algorithms are currently available in \textsc{Caesar} to extract extended sources from a suitable input map (typically a residual map):
\begin{enumerate}
\item \emph{Wavelet transform}: input map is decomposed in $J$ Wavelet scales (typically $J$=6-7) and extended sources are extracted from higher scales by thresholding (e.g. employing the same algorithm used for compact sources);  
\item \emph{Saliency filtering}: a multiscale saliency filter is applied to the input map and extended sources are extracted from the filtered image by thresholding (e.g. employing the same algorithm used for compact sources);
\item \emph{Hierarchical clustering}: input map is oversegmented into a series of superpixels (or regions) on the basis of a spatial and flux similarity measure. Neighbouring regions are then adaptively merged by mutual similarity and a final segmentation into background and sources is obtained. The method was presented in \cite{Riggi2016} and it is currently being updated to lower its computing resource demand;
\item \emph{Active contour}: the method iteratively determines the contour that best separates objects from the background in the input image, starting from an initial curve and evolving it by minimizing an energy functional until convergence. Two different algorithms are available, one based on the Chan-Vese active contour model \citep{ChanVese2001} and the other based on the localizing region-based active contour (LRAC) model \citep{Lankton2008};
\end{enumerate}

\subsubsection{Source merging}
\label{sec:sourcemerging}
Two source merging steps can be optionally included in the pipeline. The first is performed by workers at the end of each tile processing task to merge overlapping extended and compact sources found by different algorithms. This step was introduced to allow full detection of faint extended sources with compact brighter components. Indeed, the compact source finder typically detects only the bright regions, while the extended finder detects only the diffuse part, particularly if the former was removed/subtracted in the input residual map.\\A second merging step is performed by the master process after gathering all sources detected by workers. Sources located at the edge or in overlapping regions of neighbouring tiles are merged if adjacent or coincident.

\section{Simulated data}
\label{sec:simulations}
In order to test source finding performances, we generated simulated sky models (2560$\times$2560 pixels, 1" pixel size) with both point and extended sources uniformly distributed in ($\alpha$,$\delta$). A source density of 1000 deg$^{-2}$ was assumed for point-sources and 50 deg$^{-2}$ for extended sources. Source densities assumed in the simulation correspond to values measured in the \textsc{Scorpio} ATCA survey \citep{Umana2015}. Source peak brightness $S_{peak}$ was randomly generated with a uniform distribution in $\log(S_{peak})$ in the range $S_{peak}$=[0.1,1000] $\mathrm{mJy/pixel}$ for point sources and $S_{peak}$=[1,100] $\mathrm{\mu Jy/pixel}$ for extended sources. The peak brightness distribution assumed was driven by the need of having a sufficient number of simulated sources for statistical analysis over the entire flux range, rather than by physical considerations or existing observations. Extended sources were generated with equal proportion weights from five different shape models (disk+shell, ring, ellipse, gaussian, S\'{e}rsic profile) with a maximum angular scale of 10 arcmin\footnote{For gaussian and S\'{e}rsic source generation models the maximum angular scale assumed corresponds to the standard deviation and effective radius respectively.}.\\
For each sky model we simulated 12 $hr$ observations with the Australia Telescope Compact Array (ATCA) using \textsc{casa} tool \citep{McMullin2007}. All available ATCA configurations were used. 
Eight pointings were needed to cover the sky model area given the ATCA primary beam.\\
The imaging stage was performed in an automated way assuming a 100 $\mathrm{\mu Jy}$ clean threshold and cleaning mask boxes around each generated source. Simulated fields were imaged singularly and combined afterwards to produce the final simulated mosaic. To limit computing time the imaging process was not fully optimized. In fact the focus was put in achieving a sufficient imaging of both compact and extended objects to carry out source finding.
A number of 200 simulated mosaics are available to test source finding. The average noise level is 300-400 $\mathrm{\mu Jy}$ with the chosen imaging parameters and mosaic strategy. The synthesised beam of simulated maps is $b_{maj}$=13.3", $b_{min}$=8.4", $b_{pa}$=0$^{\circ}$.
Although a number of effects have been neglected or ideally modeled (e.g. perfect calibration is considered), the simulated maps include typical interferometric noise patterns and can be used as a valid test bench for existing source finders. To this aim the entire simulated dataset was made publicly available at \url{http://doi.org/10.5281/zenodo.3257594}.\\

\section{Analysis}
\label{sec:analysis}
In this section we report the detection performances for compact and extended sources obtained on the simulated data sample described in section~\ref{sec:simulations}.

\subsection{Validation of simulated data}
To test the imaging quality we compared generated and imaged sources following this approach. Each generated source was convolved with the synthesis beam and the resulting image thresholded to keep 99\% of the total source flux. The convolved source mask obtained represents the ground truth. Imaged sources are obtained by applying the convolved source mask to the simulated mosaic. Overlapping compact and extended sources were merged so that three classes of sources (point, extended + point, extended) have to be inspected.\\A sample simulated map with convolved source contours for the three classes (point-like in red, extended in blue, extended+point-like in green) is shown in Fig.~\ref{fig:imagingperf} (left panel). In Fig.~\ref{fig:imagingperf} (right panel) we compared the flux density of convolved and imaged sources (with background subtracted) for the three classes of sources and using the full simulated data set. As can be seen, fluxes are reconstructed with an accuracy better than 10\% for bright sources, increasing to 40\% for very faint sources. Systematic biases are found below 10\%.\\ 
In the following analysis we will take the imaged source flux as the reference when evaluating the flux uncertainty of the source finding process.

\subsection{Detection of compact sources}
Compact source finding was run on the $N$=200 simulated maps using the set of parameters reported in Table~\ref{tab:compactfinderparameters}. A number of 71,640 generated point-sources are available for analysis.\\Sources tagged as point-like were cross-matched in position to generated sources. A generated source is labelled as "detected" if the distance between its centroid and the one of a measured source is smaller than 10" (corresponding to 10 pixels and slightly less than the average of beam dimensions). If many measured source candidates are present, the matched source will be the one with the shortest distance. 

\begin{table}[!ht]
\caption{Compact source finder parameters.}
\centering%
\footnotesize%
\begin{tabular}{|ll|}
\hline%
\rowcolor{black}
\textcolor{white}{Parameter} & \textcolor{white}{Value} \\
\hline%
\rowcolor{silver}
\multicolumn{2}{|c|}{\emph{Bkg/Noise}}\\%
bkg & median\\%
noise & mad\\%
box size & 10$\times$beam\\%
grid step & 20\%box\\%
\hline%
\rowcolor{silver}
\multicolumn{2}{|c|}{\emph{Blob Detection}}\\%
$Z_{thr,d}$ & 5\\%
$Z_{thr,m}$ & 2.5\\%
$n_{pix}$ & 5\\%
$n_{iter}$ & 2\\%
$\Delta\sigma_{seed}$ & 0.5\\
\hline%
\rowcolor{silver}%
\multicolumn{2}{|c|}{\emph{Nested Blob Detection}}\\%
method & LoG\\%
min scale & 1$\times$beam\\
max scale & 2$\times$beam\\%
scale step & 1\\%
$Z_{thr,d}$ & 5\\%
$Z_{thr,m}$ & 2.5\\%
$n_{beams}^{thr}$ & 20\\%
\hline%
\rowcolor{silver}%
\multicolumn{2}{|c|}{\emph{Source fitting}}\\%
$n_{beams}^{thr}$ & 10\\%
max components & 5\\%
$Z_{thr,peak}$ & 1\\%
bkg offset & fixed\\%
\hline%
\end{tabular}
\label{tab:compactfinderparameters}
\end{table}

\begin{figure*}[!ht]
\begin{center}
\subtable[Position accuracy]{\includegraphics[scale=0.5]{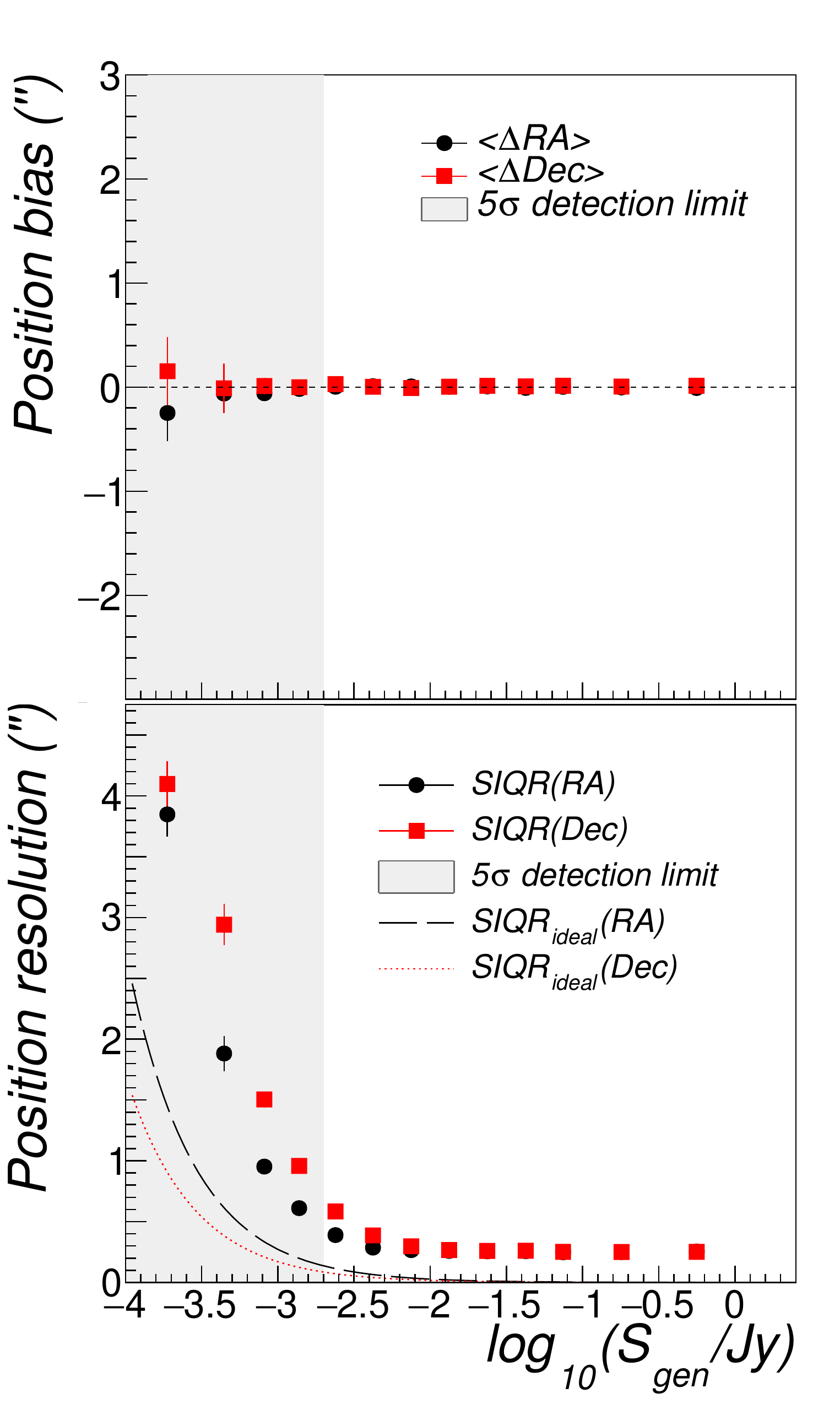}}%
\subtable[Flux density accuracy]{\includegraphics[scale=0.5]{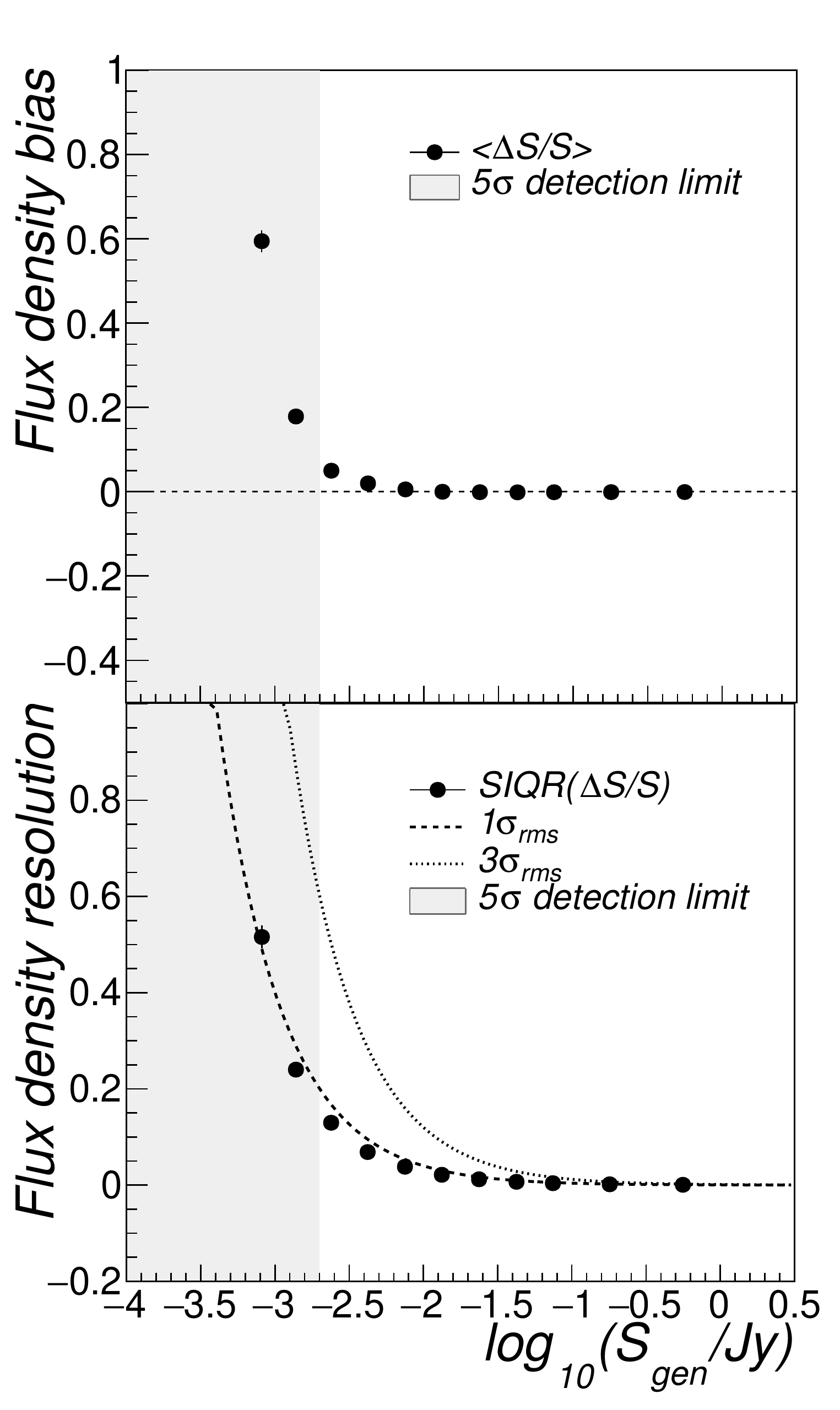}}%
\caption{Left: Compact source position reconstruction bias (upper panel) and resolution (bottom panel) as a function of the source generated flux. Bias is estimated using sample median in each flux bin, while resolution is computed using the semi-interquartile range (SIQR). Black dots and red squares indicate RA and Dec coordinates respectively. Dashed and dotted lines denote the ideal resolution in both coordinates computed with expression~\ref{formula:idealreso} (see text). Right: Compact source flux density reconstruction bias (top panel) and resolution (bottom panel) as a function of the source generated flux. Dashed and dotted lines indicate the expected 1$\sigma_{rms}$ and 3$\sigma_{rms}$ flux density errors respectively, with $\sigma_{rms}$=400 $\mu$Jy rms noise level.}
\label{fig:posfluxaccuracy}
\end{center}
\end{figure*}

\subsubsection{Completeness and reliability}
Following the described procedure we computed the source completeness (or detection efficiency) and reliability metrics for the simulated data sample. Completeness, at a given level of quality selection, denotes the fraction of generated point-sources identified by the source finder, given the assumed match criteria, and passing the imposed selection cuts.
Reliability is the fraction of detected sources passing the quality selection that corresponds to real sources.
Completeness and reliability are reported in Fig.~\ref{fig:compactsourcecompleteness} for four different selection cuts as a function of the source generated and measured flux density respectively. 
The gray shaded area indicates a region of source significance below 5$\sigma$, assuming an average rms of 400 $\mu$Jy.
Black dots (labelled as "fit") are obtained using a minimal set of quality cuts:
\begin{itemize}
\item Source match in position
\item Source fit performed and converged
\item Positive fitted amplitude parameters
\end{itemize}
Red squares (labelled as "presel") corresponds to high-quality fitted sources, passing the following preselection cuts:
\begin{itemize}
\item Fit $\chi^{2}$/ndf<10
\item Accurate fit error matrix (flag returned by fit minimizer)
\end{itemize}

Blue diamonds and green triangles correspond to two additional quality selection cuts applied to the detected sources after preselection (described below).\\
As can be seen, 90-95\% of the generated sources at a 5$\sigma$ flux significance are detected, assuming the finder parameters listed in Table~\ref{tab:compactfinderparameters} and the preselection cuts. The corresponding false detection rate is of the order of 20\% at 5$\sigma$ and 5-10\% at larger significance levels. False detections are largely due to the over-deblending of imaging artefacts and extended sources present in the simulated maps. For instance, we report in Fig.~\ref{fig:samplefalsesource} examples of false sources (shown with red contours) detected in two different simulated maps. White contours represents true point-sources, while green contours are the sources detected by the \textsc{Aegean} \citep{Hancock2018} source finder for comparison. As can be seen, faint diffuse emission induces a large number of source components in the deblending process. This effect, expected to be observed in all finders implementing a deblending stage, is apparently less evident in similar analysis reported in the literature (e.g. see \cite{Hopkins2015}). This is most likely due to a combinations of multiple factors: a better imaging in the (real or simulated) data, the absence of extended sources in the test samples, the usage of tighter quality cuts.\\%
Over-deblending can be partially prevented in \textsc{Caesar} by increasing the source significance threshold and the deblending threshold parameters (peak threshold, $n_{beams}^{thr}$ for fitting or the maximum number of fitted components). However, we have found that with a different choice of deblending parameters the reliability can be slightly increased but no more than a few percent. We therefore tried applying a further selection to the data to identify false sources. For this we have exploited the physical consideration that true fitted point-like sources are expected to be morphologically similar to the beam and thus defined three classification parameters: 
\begin{itemize}
\item $\delta\theta$: rotation angle (in degrees) of source fitted ellipse with respect to the beam ellipse, expected to be peaked around 0 for true sources, provided that the beam is elliptical in shape (as in this analysis);
\item $E_{source}/E_{beam}$: source fitted ellipse eccentricity divided by the beam ellipse eccentricity, expected to be peaked around 1 for true sources;
\item $A_{source}/A_{beam}$: source fitted ellipse area divided by the beam area, expected to be peaked around 1 for true sources;
\end{itemize}

\begin{figure*}[!ht]
\begin{center}
\subtable[Completeness]{\includegraphics[scale=0.4]{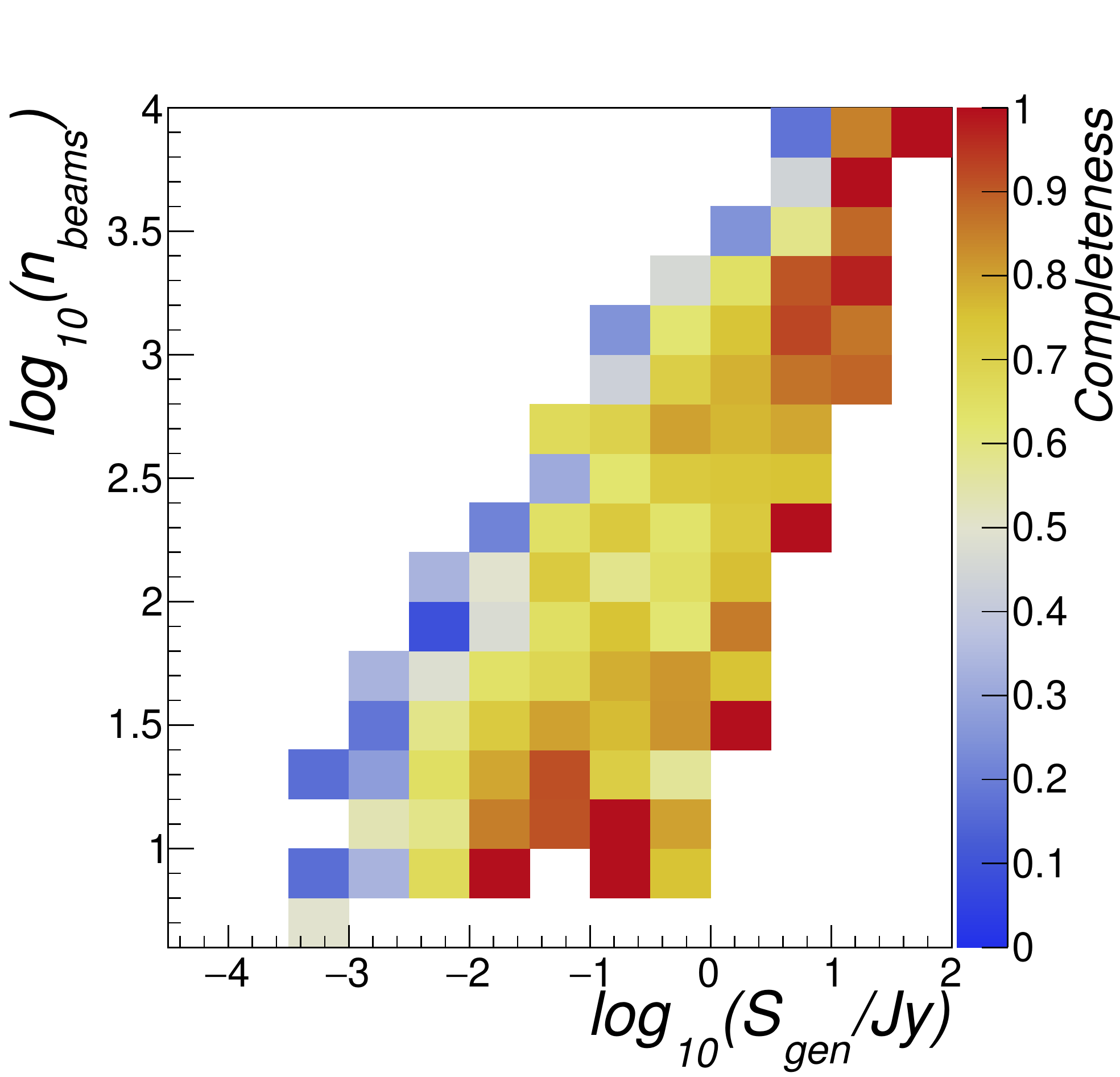}}%
\subtable[Reliability]{\includegraphics[scale=0.4]{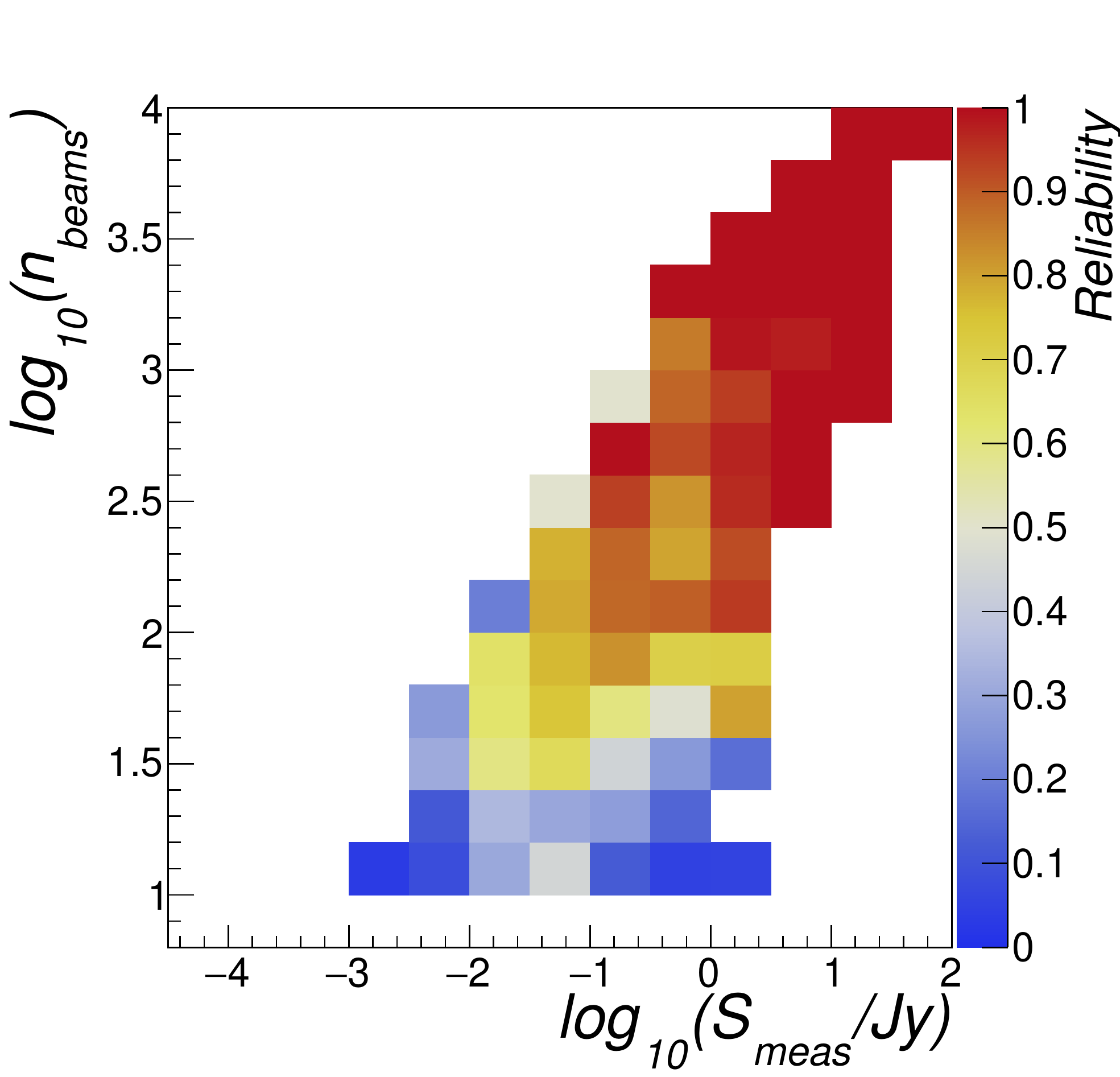}}%
\caption{Left: Extended source detection efficiency as a function of the generated source flux density and $n_{beams}$ (multiple of the synthesized beam size); Right: Extended source detection reliability as a function of the measured source flux density and $n_{beams}$.}
\label{fig:extsourcecompleteness}
\end{center}
\end{figure*}

In Fig.~\ref{fig:truefalseparhisto} we report the distributions of the three parameters for real (red histograms) and false (black histograms) sources. Using these parameters we have setup two different classifiers:
\begin{itemize}
\item \textit{Cut-based classifier}: Sources passing these quality cuts on the three source parameters are selected as real:
  \begin{itemize}
  \item $|\delta\theta|<$45
  \item 0.5<$E_{source}/E_{beam}$<1.5
  \item 0.1<$A_{source}/A_{beam}$<10
\end{itemize}
Cuts are not fine-tuned and no correlation among variables is taken into account (e.g. cuts are derived separately for each parameter). Cut values can be customized in the finder configuration file.
\item \textit{Neural network classifier}: We trained a multi-layer perceptron (MLP) neural network (NN) on 50\% of the available source sample to identify real and false sources using the three parameters as input variables\footnote{We are deliberately employing in this paper the simplest neural network architecture possible (i.e. MLP with two hidden layers) trained with only three input parameters. In the future we plan to increase performances by employing more advanced deep learning network architectures (e.g. convolutional neural networks) working on the full image pixel data.\\Moreover, additional simulated maps are planned to be generated to provide a completely independent training sample with respect to the one currently used for testing.}.
\end{itemize}
Both classifiers were applied to the full set of detected sources and completeness/reliability were computed on the selected data sample. We reported the obtained results in Fig.~\ref{fig:compactsourcecompleteness}: blue diamonds are relative to the cut-based classifier, green triangles are obtained using the NN classifier. As can be seen both classifiers allow to increase the detection reliability by $\sim$10-15\% at the cost of a moderate completeness degrade. The neural network approach, working on a joint set of classification variables and providing a non-linear decision boundary, outperforms, as expected, the simpler cut analysis.

\subsubsection{Position and flux accuracy}
We report in Fig.~\ref{fig:posfluxaccuracy} the source position (left panel) and flux density (right panel) accuracy as a function of the source generated flux obtained over preselected source sample. Reconstruction bias is estimated using the sample median in each flux bin and reported in the top panels. Statistical resolution is estimated using the semi-interquartile range and reported in the bottom panels.\\  
Ideal position resolutions in both coordinates are reported in the bottom left panel and given by (see \citep{Condon1997}):
\begin{eqnarray}\label{formula:idealreso}
IQR(x)=f\times\sqrt{\frac{2\sigma_{x}}{\pi\sigma_{y}}}\frac{\sigma_{rms}}{A}h\cr%
IQR(y)=f\times\sqrt{\frac{2\sigma_{y}}{\pi\sigma_{x}}}\frac{\sigma_{rms}}{A}h
\end{eqnarray}
with $h$=1" map pixel scale size, $A$ source peak flux, $\sigma_{x,y}$ source Gaussian sigma in the x and y directions ($b_{maj}\sim$13.3", $b_{min}\sim$8.4"), $\sigma_{rms}$=400 $\mu$Jy image noise rms, $f\sim$0.674 factor to convert from gaussian standard deviation to semi-interquartile range. Typical values are $\sim$0.2" at 5$\sigma$ and $\sim$0.05" at 20$\sigma$ source significance levels. The reconstructed position uncertainties above 5$\sigma$ are found of the order of 0.4-0.5". No significant position bias is found even well below the source detection threshold.\\Flux reconstruction presents a small positive bias ($\sim$5-10\%) near the detection threshold. A similar trend was found also in other finders \citep{Hopkins2015}. Flux accuracy is found better than few percents for bright sources and $\sim$10\% at the detection threshold.

\subsection{Detection of extended sources}
\label{sec:extsourceanalysis}
Extended source finding was run on the $N$=200 simulated maps. A number of 3459 generated extended sources are available for this analysis. For this work we considered the saliency filtering algorithm among those available in \textsc{Caesar}. The algorithm steps were summarized in Section~\ref{sec:extsourceextraction} and extensively described in \cite{Riggi2016}. 
Algorithm parameters used in this analysis are reported in Table~\ref{tab:extfinderparameters}.\\ 
Sources tagged as extended or extended+compact were cross-matched to generated sources (convolved with the synthesis beam as described in Section~\ref{sec:simulations}) using overlap area and flux ratio parameters. 

\begin{table}[!ht]
\caption{Extended source finder parameters.}
\centering%
\footnotesize%
\begin{tabular}{|ll|}
\hline%
\rowcolor{black}
\textcolor{white}{Parameter} & \textcolor{white}{Value} \\
\hline%
\rowcolor{silver}
\multicolumn{2}{|c|}{\emph{Compact source filter}}\\%
$Z_{thr,low}^{res}$ & 5\\%
$Z_{thr,high}^{res}$ & 10\\%
filter kernel size (\#pix) & 21\\
removed sources & point-like\\
\hline%
\rowcolor{silver}
\multicolumn{2}{|c|}{\emph{Smoothing filter}}\\%
filter model & guided\\%
radius (\# pix) & 12\\%
$\varepsilon$ & 0.04\\%
\hline%
\rowcolor{silver}
\multicolumn{2}{|c|}{\emph{Saliency filter}}\\%
spSize (\# pix) & 20\\%
spBeta & 1\\%
spMinArea & 10\\%
saliencyResoMin (\#pix) & 20\\%
saliencyResoMax (\#pix) & 60\\%
saliencyResoStep (\#pix) & 20\\%
saliencyNNFactor & 1\\%
saliencyThrFactor & 2.5\\%
\hline%
\end{tabular}
\label{tab:extfinderparameters}
\end{table}

A generated source $i$ is considered as "detected" by a measured source $j$ if: 
\begin{itemize}
\item $n_{i\cap j}/n_{i}>f_{thr}^{high}$
\item $n_{i\cap j}/n_{j}>f_{thr}^{high}$
\end{itemize}
or, alternatively, if:
\begin{itemize}
\item $n_{i\cap j}/n_{i}>f_{thr}^{low}$,\;$t_{thr}^{min}<S_{i\cap j}/S_{i}<t_{thr}^{max}$
\item $n_{i\cap j}/n_{j}>f_{thr}^{low}$,\;$t_{thr}^{min}<S_{i\cap j}/S_{j}<t_{thr}^{max}$
\end{itemize}
where:
\begin{itemize}
\item[-] $n_{i}$: number of pixels in generated source $i$
\item[-] $n_{j}$: number of pixels in measured source $j$
\item[-] $n_{i\cap j}$: number of overlapping pixels between generated and detected sources
\item[-] $S_{i}$: sum of pixel fluxes for generated source $i$
\item[-] $S_{j}$: sum of pixel fluxes for measured source $j$
\item[-] $S_{i\cap j}$: sum of pixel fluxes for measured source $j$, computed over pixels overlapping with generated source $i$
\end{itemize}
$f_{thr}^{high}$, $f_{thr}^{low}$, $t_{thr}^{min}$, $t_{thr}^{max}$ are configurable thresholds, assumed equal to 60\%, 10\%, 80\% and 120\% respectively in this work. 
The first condition imposes a large overlap area between generated and measured sources without any condition applied on their fluxes. The second condition, instead, requires a minimal overlap area plus a high match between fluxes. 

\begin{figure}[!ht]
\begin{center}
\includegraphics[scale=0.45]{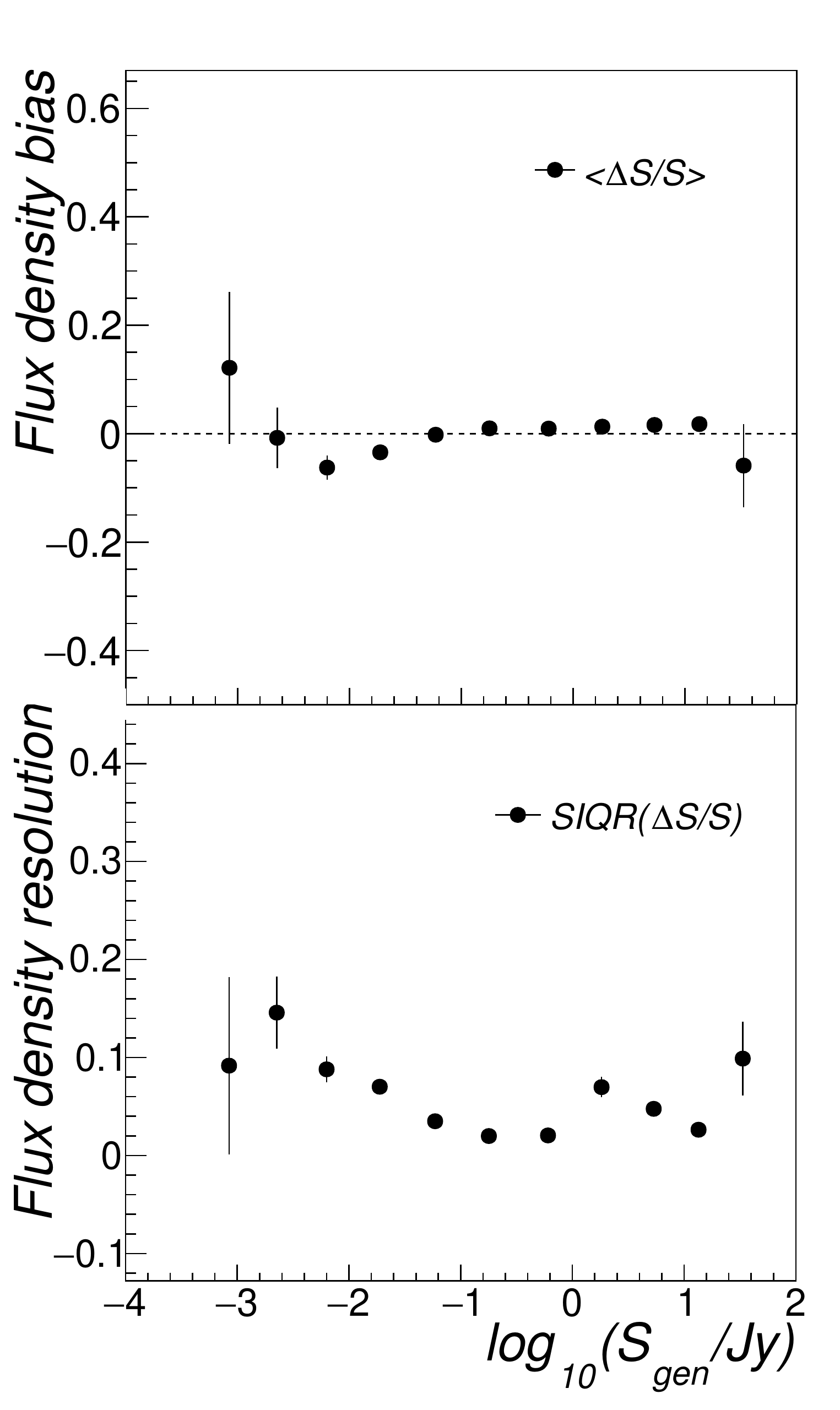}
\caption{Extended source flux density reconstruction bias (top panel) and resolution (bottom panel) as a function of the source generated flux. Bias is estimated using sample median in each flux bin, while resolution is computed using the semi-interquartile range (SIQR).}
\label{fig:extsourceposfluxaccuracy}
\end{center}
\end{figure}

\subsubsection{Completeness and reliability}
Similarly to what has been done for compact sources, we computed the completeness and reliability obtained for extended sources as a function of generated/measured source flux density and $n_{beams}$ (multiple of the synthesized beam size). Results are reported in Fig.~\ref{fig:extsourcecompleteness}. Completeness is on average 60\%-70\% for fainter sources and $\sim$80\% for brightest sources. Reliability is found of the order of $\sim$70\% on average and above 90\% for high flux densities. Detection efficiency slightly degrades for sources with size comparable with the minimum spatial scale assumed in the finding algorithm. A similar trend is observed for the largest sources injected in the simulation. For a given flux density this is due to their intrinsic smaller pixel detection significance.\\
Given the limited size of the simulated sample currently available we are not able to disentangle the relative contributions of different simulated source types (ring, disk + shell, ellipse, S\'{e}rsic, gaussian) in the above trends. 

\begin{figure*}[!ht]
\centering%
\subtable[Multithread speedup]{%
  \includegraphics[scale=0.4]{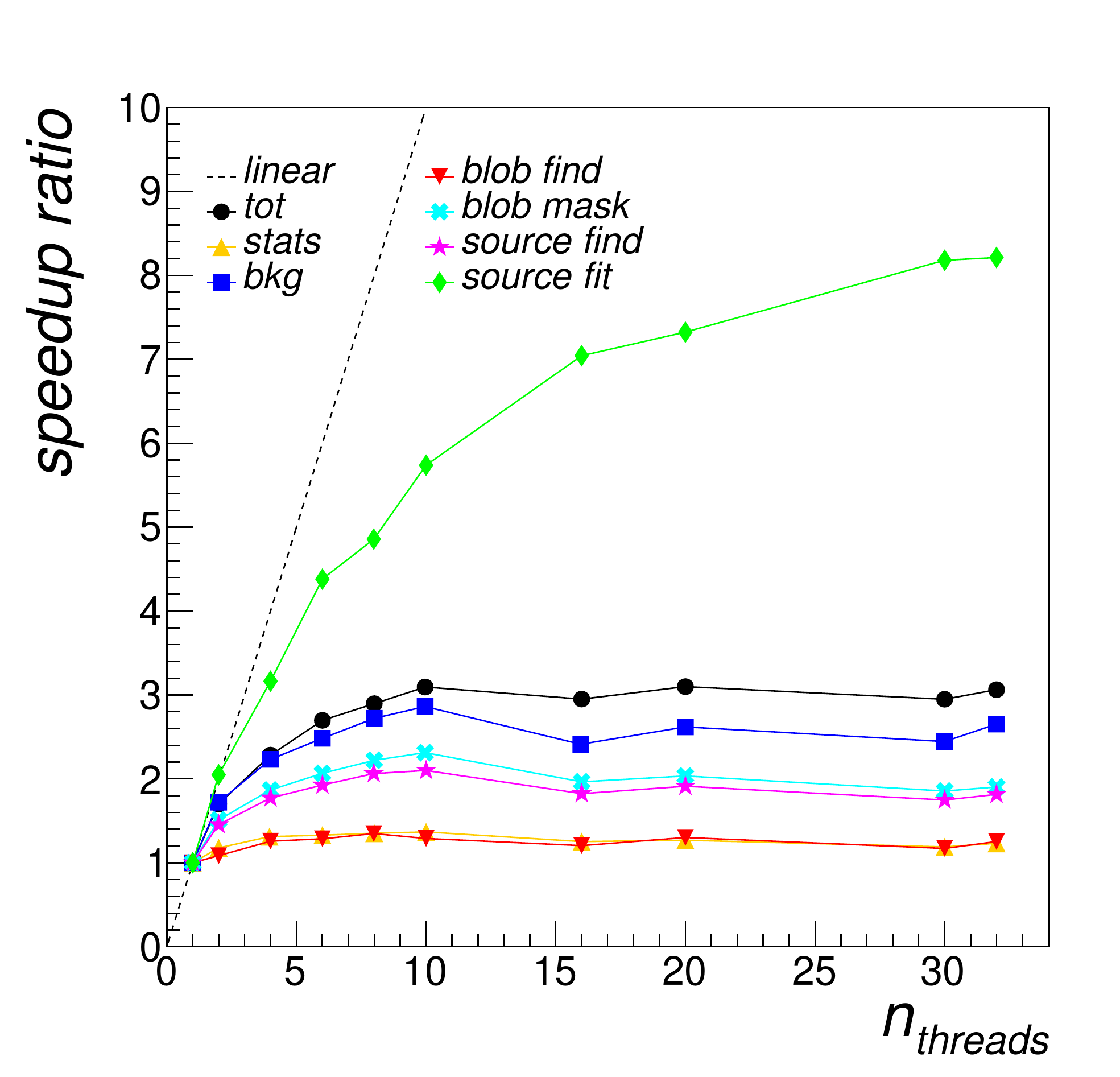}
}%
\subtable[CPU time fraction]{%
  \includegraphics[scale=0.4]{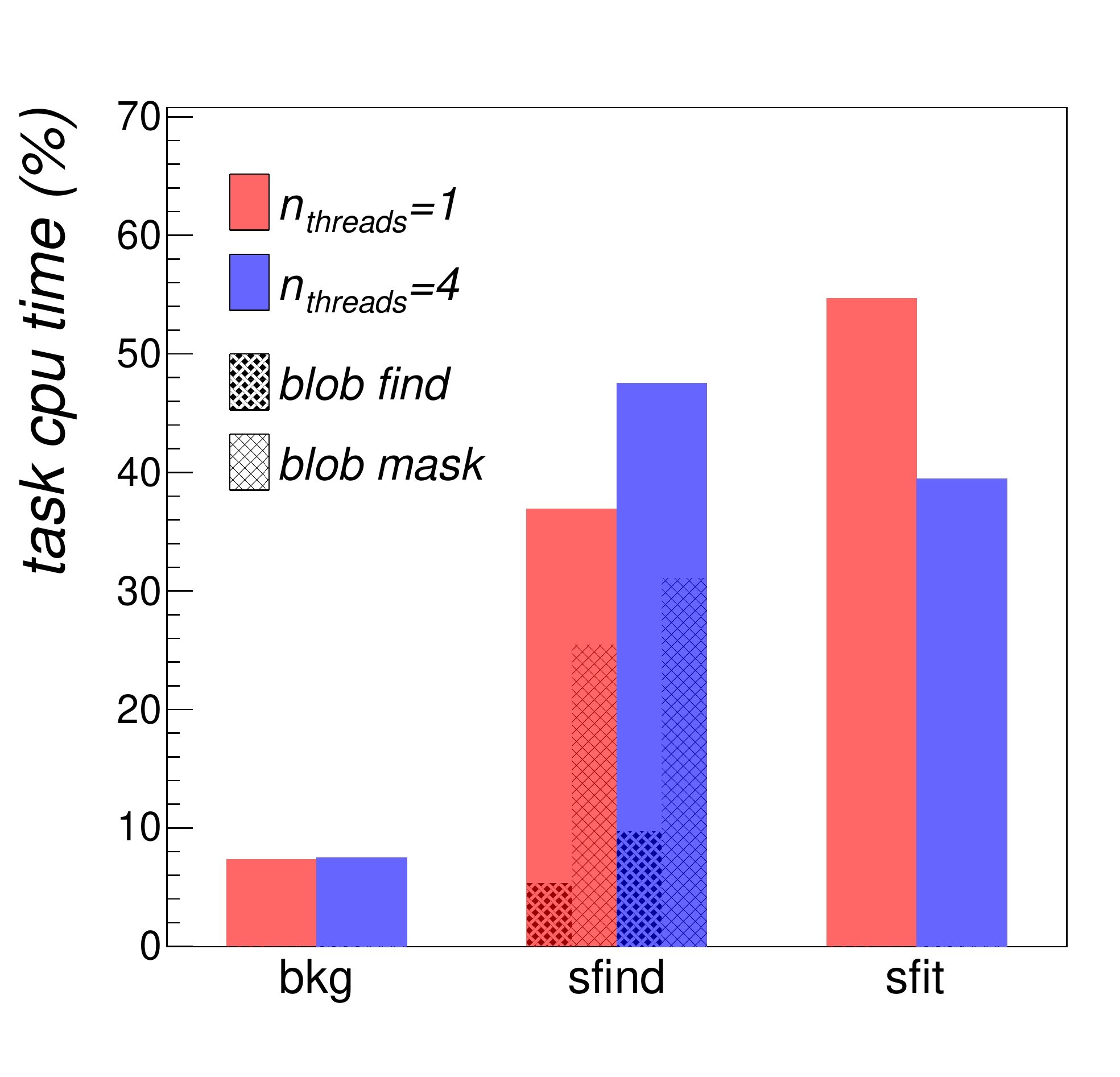}
}%
\caption{Left: Computational speedup of multithreaded compact source finding over a 10000$\times$10000 pixel map as a function of the number of allocated threads (black line) compared with the ideal speedup (black dashed line). Colored lines indicate the speedup obtained on different tasks: image statistic calculation (orange line), image background calculation (blue line), source finding (purple line), source fitting (green line). Source finding is further decomposed in two subtasks: blob finding (red line), blob mask (light blue line). Right}: Fraction of the total cpu time spent in different source finding tasks with $n_{threads}$=1 (red histogram) and $n_{threads}$=4 (blue histogram).%
\label{fig:multithreadperf}
\end{figure*}

We therefore limit our report below (see Table~\ref{tab:extsourceclasscompleteness}) to the detection efficiency obtained for different source classes irrespective of flux density and source size. Sources formed by a combination of different types have been labeled as "mixed". Sources of a given pure class having point-like sources inside were still considered as belonging to the same class.

\begin{table}[!ht]
\caption{Detection efficiency $\varepsilon$ for different extended source types.}
\centering%
\footnotesize%
\begin{tabular}{ll}
\hline%
\rowcolor{silver}
\textcolor{black}{Source type} & \textcolor{black}{$\varepsilon$ (\%)} \\
\hline%
ring & 58\\ 
disk+shell & 81\\ 
ellipse & 82\\ 
S\'{e}rsic & 61\\ 
gaussian & 69\\ 
mixed & 80\\%
\hline%
\end{tabular}
\label{tab:extsourceclasscompleteness}
\end{table}

These results suggest that ring-shaped sources and sources with tailed flux profiles (S\'{e}rsic, gaussian) are harder to be identified with respect to other types.\\As expected, the detection performances are not at the same level of point sources. Nevertheless, as we have shown in \cite{Riggi2016} with real interferometric data, the outcomes would have been considerably worse or even close to a null detection efficiency if we had used the same algorithm used for compact source. 

\begin{figure}[!ht]
\centering%
\includegraphics[scale=0.4]{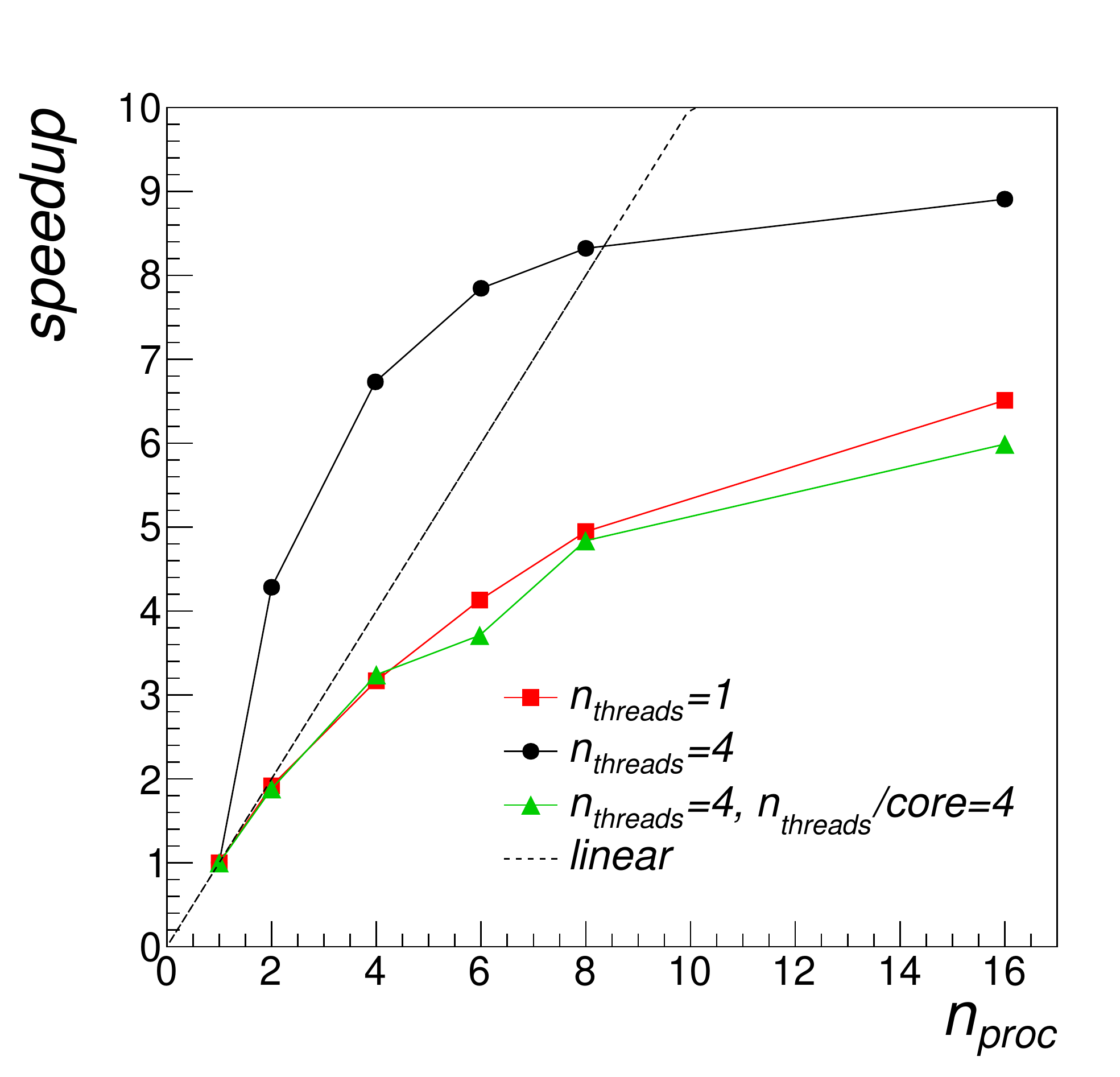}%
\caption{Computational speedup of compact source finding in MPI+OpenMP runs over a 32000$\times$32000 pixel simulated map as a function of the number of allocated MPI processes using $n_{threads}$=1 (red squares) and $n_{threads}$=4 (black dots) per MPI process. Green triangles refer to the speedup obtained using $n_{threads}$=4 per MPI process, with all four threads running on the same computing core rather that in a dedicated core.}
\label{fig:mpiperf}
\end{figure}

\subsubsection{Flux accuracy}
In Fig.~\ref{fig:extsourceposfluxaccuracy} we report the flux accuracy (bias and resolution) obtained for the detected extended sources. Flux density was computed using the sum of pixel fluxes divided by the beam area. A flux resolution below 10\% was obtained on the selected source sample. No significant biases were found.

\section{Computational performance}
\label{sec:performances}
As discussed in Section~\ref{sec:caesardev}, \caesar was improved to support parallel processing. An hybrid programming model with two levels of parallelism was adopted. The outer MPI-based level enables to distribute source finding over image tiles on multiple processors in the available computing nodes. The inner OpenMP-based level distributes source finding tasks for a single image tile across multiple threads.\\To validate the current implementation and estimate the achieved performances we measured the computing times of compact source finding on large simulated images over this computing infrastructure:
\begin{itemize}
\item 2 computing nodes connected through a 10 Gbit network link
\item 4 sockets$\times$10 Core Intel(R) Xeon(R) CPU E5-4627 2.60 GHz per node
\item 256 GB DDR4 2133 MHz memory per node
\end{itemize}
In Fig.~\ref{fig:multithreadperf} (left panel) we report the obtained speedup as a function of the number of threads allocated by OpenMP for different source finding stages and overall (shown in black) over a simulated image of size 10000$\times$10000 pixels. In the performed runs the MPI processing was switched off and the image was not partitioned in tiles. We also imposed thread affinity on the basis of the Non-Uniform Memory Access (NUMA) architecture reported by the two computing nodes, e.g. we bound threads to run on the same socket if fitting the available number of cores per socket. As can be seen, a computational speedup $\sim$3-4 is achieved overall up to a moderate number of threads (4-6), above which no significant improvement is observed for most of the tasks. Some tasks (e.g. blob finding, background calculation, source fitting) exhibit a better scalability (up to $\sim$10 allocated threads) due to their embarrassingly parallel nature and implementation. Others (e.g. image statistic calculation, blob masking) are rather flat in speedup, either because dominated by serial parts or because affected by thread management overhead (creation, synchronization, etc).\\In Fig.~\ref{fig:multithreadperf} (right panel) we report the percentage of total cpu time spent in different finding stage for two representative number of allocated threads: 1 (red histogram), 4 (blue histogram). A cpu time of $\sim$1.3/2.6 hours was spent in total with/without splitting the input image into tiles, improving by a factor of 3-4 using 4-6 threads. 
As expected, the largest contribution is due to source finding and fitting stages. Local background and rms map calculation contributes to less than 10\% of the total cpu time. Image reading and computation of statistical estimators contribute to less than 1\%.\\
In Fig.~\ref{fig:mpiperf} we report the speedup obtained on a sample simulated image of size 32000$\times$32000 pixels as a function of the number of MPI processes used. Runs were performed on a NFS filesystem mounted on both nodes. Input image was split into tiles of size 4000$\times$4000 pixels. Red squares and black dots correspond to runs performed with 1 and 4 OpenMP threads per MPI process respectively. Green triangles refer to the speedup obtained using $n_{threads}$=4 per MPI process, with all four threads running on the same computing core rather that in a dedicated core. As can be seen a good speed up is found using one single OpenMP thread. The speedup with four OpenMP threads is superlinear up to $\sim$8 processes, above which the effect of inter-process communication and data serialization becomes dominant. This is expected given that the load is partitioned over a larger number of cores, with respect to the case $n_{threads}$=1. However, when running OpenMP threads in a single core (green triangles) rather than in a separate one (black dots), the obtained speedup is comparable to the single thread speedup (red squares).\\
The performance degrade due to the network filesystem and log activity was investigated by comparing the computing times obtained when running on a local filesystem and disabling logging. We observed an increase of $\sim$5\% in the total computing time in the NFS filesystem. Tests will be performed in the future to evaluate the benefits of using a parallel filesystem such as Lustre\footnote{http://lustre.org/} or BeeGFS\footnote{https://www.beegfs.io/content/}. Logging was also found to negatively impact performances with an increase of $\sim$40\% in the total computing time.

\section{Summary and outlooks}
\label{sec:summary}
We have presented in the paper the current status of \caesar source finder. Considerable improvements were done since the first reference paper \citep{Riggi2016}, among them distributed source processing and algorithm improvements on compact sources.\\We reported the performances achieved on both compact and extended source detection using simulated data. Results obtained on compact sources are comparable to similar analysis reported in the literature \citep{Westmeier2012,Hopkins2015}, despite the presence of background emission from extended sources (typically not included in other analysis). We discussed also possible methods to discover and remove false source detections from the final catalogue.
\\To the best of our knowledge this paper reports also a first attempt to systematically test extraction of extended sources with different shapes, other than the standard gaussian model used in other analysis. 
The overall performance achieved by the extended finder algorithm tested in this paper does not compete yet with those obtained by compact finder algorithms on point sources. Nevertheless, when considering the complete sample of sources (compact plus extended) present in the observed field, the results are encouraging since the combination of different algorithms in \textsc{Caesar} allows to recover a significant fraction of sources that would have been undetected if only using the compact source finder.\\%
Despite the progress made there is still room to extend and improve \caesar both at the code and algorithmic level. For future releases we foresee additional refinements and optimizations in the code to improve memory usage, scalability and fault tolerance of the parallel implementation. In a shorter time scale, scalability can be slightly improved by exploiting parallelism on selected tasks of the pipeline that are still serially performed, particularly in extended source extraction. In a longer term, following the current trends in exascale computing, we expect a potential boost in performances if additional developments will be made to fully exploit new generations of HPC systems equipped with high-capacity memories and one or more accelerators (GPUs or FPGAs) per node.\\The obtained results highlighted that additional efforts are to be spent to improve source finding performances in view of future large area surveys. 
For compact sources we expect that improving the deblending stage and the spurious source identification will be the major area of investigation using deep neural networks trained to identify real and false components in extracted sources.  
Extended source finding will instead require different and more refined algorithms to be tested. For this purpose, additional test campaigns are planned to be performed using the algorithms already implemented in \caesar.

\begin{acknowledgements}
We acknowledge the computing centre of INAF - Osservatorio Astrofisico di Catania, under the coordination of the CHIPP project, for the availability of computing resources and support.\\
The research leading to these results has received funding from the European Commissions Horizon 2020 research and innovation programme under the grant agreement No. 731016 (AENEAS).\\
We thank the authors of the following software tools and libraries that have been extensively used for data analysis and visualization: \textsc{Casa}, astropy, \textsc{Root}, ds9, APLpy.   
The authors thank in particular the anonymous referee for helpful comments that led to the improvement of this paper.
\end{acknowledgements}

\begin{appendix}
\section{Source deblending and fit initialization}
\label{appendix:deblending}
The number of components to be fitted for a detected source is set to the number of nested blobs, eventually ordered and selected by significance level and limited to a maximum number (5 for example). Starting values for component fit parameters are determined from blob moments.\\If no nested blobs are present in the source, the number of components and relative starting parameters are estimated with the following algorithm:
\begin{enumerate}
\item Compute blob masks at different configurable scales (usually 1 to 3 times the beam size). A blob mask is obtained by thresholding the source image convolved by a LoG kernel at a given scale.
\item Find peaks in blob masks with a dilation filter using different kernel sizes (3, 5, 7 pixels by default).
\item Reject peaks below desired flux significance level.
\item Compare surviving peaks found at different scales. If multiple peaks match within a tolerance (1-2 pixels usually) consider the one with the largest intensity and select the blob optimal scale.
\item Set number of estimated components to selected peaks, again ordered by significance level and limited up to a maximum number.
\item Set initial fit component centroid and amplitude to the peak position and flux respectively.
\item Estimate fit component shape from the previously computed masks at optimal blob scale using a Watershed segmentation algorithm seeded to the detected peak. Compute initial fit component sigma and position angle parameters from segmented blob moments. Fallback to beam parameters if segmentation fails.
\end{enumerate}
The starting offset parameter can be either specified by the user from the configuration file or determined from the map, e.g. set to the estimated background averaged over source pixels or computed in a box centered on the source. 
If desired, the offset parameter can be included as a free parameter in the fit. By default, however, it is kept fixed as pixel data included in the fit (down to 2.5$\sigma$ significance) do not allow the possibility to fully constrain it.

\end{appendix}

\end{document}